\documentclass{article}

\usepackage[preprint,nonatbib]{neurips_2024}
\usepackage{booktabs}

\usepackage{subcaption}
\usepackage[utf8]{inputenc} % allow utf-8 input
\usepackage[T1]{fontenc}    % use 8-bit T1 fonts
\usepackage{hyperref}       % hyperlinks
\usepackage{url}            % simple URL typesetting
\usepackage{booktabs}       % professional-quality tables
\usepackage{amsfonts}       % blackboard math symbols
\usepackage{nicefrac}       % compact symbols for 1/2, etc.
\usepackage{microtype}      % microtypography
\usepackage{xcolor}         % colors
\usepackage{algorithm}%
\usepackage{algorithmicx}%
\usepackage{algpseudocode}%
\usepackage{graphicx}
\usepackage{svg}
\usepackage{amsmath}
\usepackage{amssymb}

\title{I$^2$VC: A Unified Framework for \\
 Intra- \& Inter-frame Video Compression}

\author{
  \hspace{-1cm}Meiqin Liu\thanks{\;Equal contribution} \\
  \hspace{-1cm}Institute of Information Science \\
  \hspace{-1cm}Beijing Jiaotong University \\
  \hspace{-1cm}Beijing, China 100044 \\
  \hspace{-1cm}\texttt{mqliu@bjtu.edu.cn} 
  \And
 Chenming Xu$^{*}$ \\
Institute of Information Science\\
Beijing Jiaotong University\\
  Beijing, China 100044 \\
  \texttt{chenming\_xu@bjtu.edu.cn} 
\And 
YuKai Gu$^{*}$ \\
Institute of Information Science \\
Beijing Jiaotong University \\
Beijing, China 100044 \\
\texttt{yukai.gu@bjtu.edu.cn}
  \AND
  Chao Yao\thanks{\, Corresponding author} \\
  School of Computer and Communication Engineering \\ % 
  University of Science and Technology Beijing \\
  Beijing, China 100083\\
  \texttt{yaochao1986@gmail.com} \\
  \And
  Yao Zhao \\
  Institute of Information Science \\
  Beijing Jiaotong University \\
  Beijing, China 100044 \\
  \texttt{yzhao@bjtu.edu.cn} \\
}

\begin{document}

\maketitle

\begin{abstract}
Video compression aims to reconstruct seamless frames by encoding the motion and residual information from existing frames. Previous neural video compression methods necessitate distinct codecs for three types of frames (I-frame, P-frame and B-frame), which hinders a unified approach and generalization across different video contexts. Intra-codec techniques lack the advanced Motion Estimation and Motion Compensation (MEMC) found in inter-codec, leading to fragmented frameworks lacking uniformity. Our proposed \textbf{Intra- \& Inter-frame Video Compression (I$^2$VC)} framework employs a single spatio-temporal codec that guides feature compression rates according to content importance. This unified codec transforms the dependence across frames into a conditional coding scheme, thus integrating intra- and inter-frame compression into one cohesive strategy. Given the absence of explicit motion data, achieving competent inter-frame compression with only a conditional codec poses a challenge. To resolve this, our approach includes an implicit inter-frame alignment mechanism. With the pre-trained diffusion denoising process, the utilization of a diffusion-inverted reference feature rather than random noise supports the initial compression state. This process allows for selective denoising of motion-rich regions based on decoded features, facilitating accurate alignment without the need for MEMC. Our experimental findings, across various compression configurations (AI, LD and RA) and frame types, prove that I$^2$VC outperforms the state-of-the-art perceptual learned codecs. Impressively, it exhibits a 58.4\% enhancement in perceptual reconstruction performance when benchmarked against the H.266/VVC standard (VTM). Official implementation can be found at \href{https://github.com/GYukai/I2VC}{https://github.com/GYukai/I2VC}
\end{abstract}
\newpage
\section{Introduction}
\label{sec1}

Video codec is designed to achieve high-quality reconstruction with the available transmission requirements. ITU/MPEG video coding standards, such as AVC/H.264~\cite{2003h264}, HEVC/H.265~\cite{2012h265}, VVC/H.266~\cite{2021h266}, incorporate three configurations, including All Intra (AI), Low Delay (LD), and Random Access (RA), to adapt to different intra- and inter-frame dependencies with three types of frames (I-frame, P-frame and B-frame), respectively. The current neural video compression methods~\cite{balle2018variational,lu2019dvc,yang2021learning} also provide different frameworks and some methods~\cite{feng2021vlvc,ladune2022aivc} provide general optimization strategies for corresponding structures, achieving excellent compression performance.

\begin{figure}[!t]
  \centering
  \includegraphics[width=\linewidth]{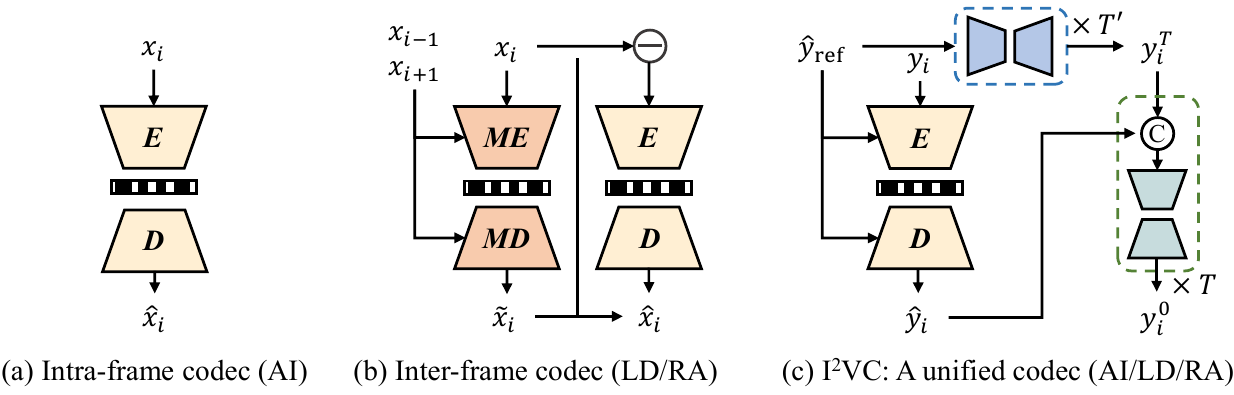}
  \caption{A comparison of different video compression frameworks. I$^2$VC employs the reference feature $\hat{y}_\text{ref}$ as a prior to compress the target feature $y_i$ and applies diffusion inversion to generate the diffusion start feature $y_i^T$, thereby incorporating inter-frame correlations.}
  \label{fig:in1}
\end{figure}

Specifically, as depicted in Figure~\ref{fig:in1}(a) and Figure~\ref{fig:in1}(b), due to the additional motion codec of inter-frame compression, different compression frameworks~\cite{balle2018variational,lu2019dvc,yang2020Learning} are designed to cater to specific frame types. It is impossible to share frameworks and weights between intra- and inter-frame compression, as well as the B-frame and P-frame of inter-frame compression, leading to model redundancy and weak generalization across existing frameworks. Therefore, it is necessary to propose a framework to unify intra- and inter-frame compression. The unified framework necessitates coherence in the compression process across three types of frames with shared certain parameter weights. Ultimately, flexible modeling of the three types of frames allows a unified model to adapt three compression configurations in a Group of Pictures (GoP). Some methods~\cite{harvey2022flexible,voleti2022mcvd,hu2023videocontrolnet} integrate inter-frame correlations directly by using reference frames or multi-modal information as a condition of the diffusion models~\cite{ho2020DDPM,song2020DDIM,rombach2022ldm,zhang2023controlnet}. However, these methods fail to address the challenge of a unified framework, specifically how to leverage inter-frame information to reduce bit-rate during inter-frame compression without Motion Estimation and Motion Compensation (MEMC). For detailed analysis and motivation, please refer to Appendix~\ref{pm} and Appendix~\ref{relate}.

% the pre-trained ControlNet is commonly used for image tasks and can not provide inter-frame correlations, leading to reconstructed videos of some methods~\cite{hu2023videocontrolnet,peng2023conditionvideo} deviating from the original texture style.

In this study, we introduce I$^2$VC, a novel framework designed to uniformly address the compression of three types of frames (I-frame, P-frame and B-frame) across different video compression scenarios (AI, LD and RA). Unlike traditional methods that rely on a motion codec for inter-frame compression, I$^2$VC achieves an integration of intra- and inter-frame compression by reference-based coding and iterative generation, as illustrated in Figure~\ref{fig:in1}(c). Specifically, a Spatio-Temporal Variable-rate Codec (STVC) and an Implicit Inter-frame Feature Alignment (IIFA) module are designed. STVC, in particular, uses the spatio-temporal significance of reference features to guide the compression of each frame type, enhancing the model's generalization. Due to without a motion codec, I$^2$VC employs a Latent Diffusion Model (LDM)~\cite{rombach2022ldm} for subtle inter-frame alignment. Notably, for inter-frame compression, the reference feature undergoes a Denoising Diffusion Implicit Model (DDIM)~\cite{song2020DDIM} inversion, setting the stage for targeted denoising of motion areas based on decoded features, effectively achieving unified compression without MEMC. Comparative experiments indicate that I$^2$VC surpasses VTM-19.0~\cite{vtm} by an impressive 58.4\% in perceptual construction performance across all configurations. In summary, our main contributions are listed as follows. 

\begin{itemize}
\item We introduce I$^2$VC, a unified framework for Intra- and Inter-frame video compression. The three types of frames (I-frame, P-frame and B-frame) across different video compression configurations (AI, LD and RA) within a GoP are uniformly solved by one framework.

\item We design a conditional coding scheme codec for three types of frames, leveraging the spatio-temporal significance of reference features to unify intra- and inter-frame correlations at variable rates. 

\item We use DDIM inversion on reference features to selectively denoise motion-rich areas, ensuring temporal consistency across frames through implicit feature alignment, bypassing the need for MEMC.

\end{itemize}

\section{Related Work}
\label{sec2}

Learned video compression adopts a similar framework to traditional compression, still requiring corresponding networks for three types of frames (I-frame, P-frame and B-frame). Towards learned I-frame compression, Ballé \emph{et al.}~\cite{balle2017end, balle2018variational} introduce a Variational Auto-Encoder (VAE) incorporating a factorized and hyperprior entropy models. Cheng \emph{et al.}~\cite{cheng2020learned} explore a Gaussian mixture entropy model to improve rate-distortion performance. Wang \emph{et al.}~\cite{wang2023evc} propose a dual spatial prior checkerboard context model to improve the probability estimation. Jiang \emph{et al.}~\cite{jiang2023mlic} propose a multi-reference entropy model to achieve state-of-the-art performance in I-frame coding. Recently, the development of generative models has improved perceptual coding performance. Mentzer \emph{et al.}~\cite{mentzer2020high} firstly utilize Generative Adversarial Network (GAN) for I-frame coding. Some methods~\cite{yang2023cdc,relic2024lossy} leverage the diffusion process to optimize the qualitative reconstruction quality of images in the decoder. Careil \emph{et al.}~\cite{careil2023towards} utilize pre-trained diffusion models to improve the generalization for low-bit-rate I-frame coding. Ma \emph{et al.}~\cite{ma2024correcting} and Muckley \emph{et al.}~\cite{muckley2023improving} respectively propose a privileged end-to-end decoder and a non-binary discriminator to improve statistical fidelity of perceptual image compression models.

Motivated by the advancement in the I-frame codec, P-frame compression methods including motion codec and residual codec are initially proposed by Lu \emph{et al.}~\cite{hu2021fvc,lu2019dvc}. For robust inter-frame prediction, Li \emph{et al.}~\cite{li2021deep} propose a Deep Contextual Video Compression (DCVC) to shift the paradigm from predictive coding to conditional coding. Sheng \emph{et al.}~\cite{sheng2022temporal} investigate Temporal Context Mining (DCVC-TCM) for P-frame compression enhancement. Moreover, Li \emph{et al.} propose a series of variable-rate video coding methods, including DCVC-HEM~\cite{li2022hybrid}, DCVC-DC~\cite{li2023dcvcdc}, and DCVC-FM~\cite{li2024dcvcfm}, which continuously optimize the performance and efficiency of the spatio-temporal checkerboard context model in the hyperprior entropy model, greatly improving rate-distortion performance and the variable range of bit-rate for P-frame compression. To exploit bi-directional correlations, some methods directly extend P-frame codecs to B-frame by using bi-directional MEMC~\cite{ccetin2022flexible,yang2021learning,yilmaz2021end} or video frame interpolation~\cite{alexandre2023hierarchical,xu2024ibvc,yang2022Advancing}. The utilization of bi-directional references renders B-frame compression frameworks incompatible with P-frame. Besides, generative networks are utilized to achieve low bit-rate video compression. Yang \emph{et al.}~\cite{yang2022perceptual} employ a recurrent conditional GAN to achieve state-of-the-art rate-perception for P-frame compression. Li \emph{et al.}~\cite{li2024extreme} achieve extreme video compression through pre-trained diffusion-based predictive frame generation. 

% fusion feature $\hat{y}_i$ is used for the controllable condition $\tau_\theta$ during the state transition process. Under limited bitrate, semantic information in the features is reconstructed through a generative approach.
 % As a result, they cannot be applied to video compression, which requires complex constraints and high inter-frame reference requirements.
 
\section{Methodology}
\label{sec3}

The proposed framework aims to adapt to the three types of frames with a unified framework, and further achieve Intra- and Inter-frame video compression. Due to the absence of motion codec in inter-frame compression for unification, using only a codec to introduce inter-frame correlations poses a challenge. We propose to use inter-frame reference features in a conditional form to reconstruct the current frame and reduce the bit-rate. The details are elaborated in the following section.

\subsection{A Unified Framework for Video Compression}
\label{VC3}

\begin{figure}[!t]
  \centering
  \includegraphics[width=\linewidth]{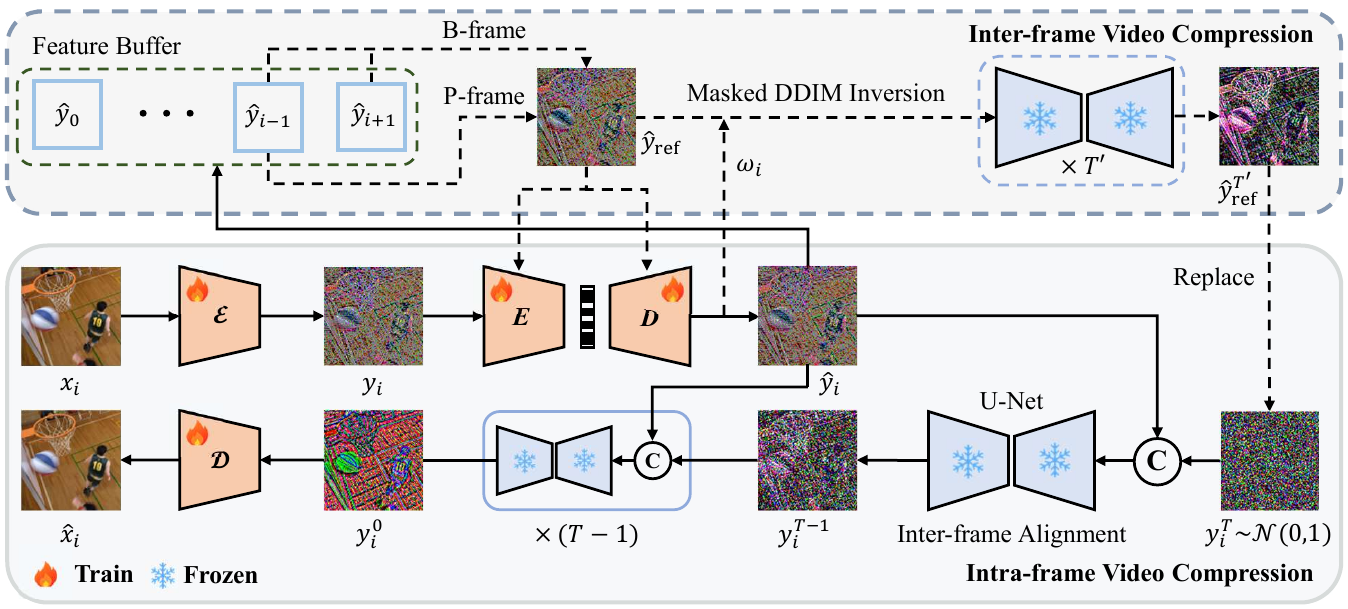}
  \caption{Overview of the unified framework for Intra- \& Inter-frame of Video Compression (I$^2$VC). The solid box represents the intra-frame compression framework, while the dashed box represents the additional reference information required for inter-frame compression. The input frame $x_i$ is encoded to fusion feature $\hat{y}_i$ and is referred to $\hat{y}_\text{ref}$ in inter-frame compression. The random Gaussian noise $y_i^T$ is transited to $y_i^0$ conditioned on $\hat{y}_i$ in $T$ denoising steps. In inter-frame compression, the masked diffusion inversion on reference feature $\hat{y}_\text{ref}$ is employed as the initial state $y_i^T$ to achieve implicit alignment. The output frame $\hat{x}_i$ is up-sampled by the pre-trained LDM decoder.}
  \label{me1}
\end{figure}

Given an input video sequence $\left\{x_i {\mid} i \in\{0,1,\ldots,n\}\right\}$ consisting of $n$ frames, video compression aims to reconstruct the high-quality video sequence $\{\hat{x}_i {\mid} i\in\{0,1,\ldots,n\}\}$ while maintaining a high compression ratio. To meet the above requirements, a unified framework based on controllable condition diffusion for Intra- \& Inter-frame Video Compression, I$^2$VC is proposed. As illustrated in Figure~\ref{me1} and Algorithm~\ref{algo1} (Appendix~\ref{al}), the input frame $x_i$ is first transformed by 4 $\times$ down-sampling convolution $\mathcal{E}(\cdot)$ to latent $y_i$ for complexity reduction. Then, the codec uses the reference features to guide the compression of each frame type. Towards different types of frames, the input frame feature $y_i$ is encoded to feature $\hat{y}_i$ with the reference feature $\hat{y}_\text{ref}$ as the prior, formulated as:
\begin{equation}
\hat{y}_i=D(\left\lfloor E(y_i, \hat{y}_{\text{ref}})\right\rceil, \hat{y}_{\text{ref}}),
\label{concat}
\end{equation}
where the encoder $E(\cdot)$ and decoder $D(\cdot)$ means the spatio-temporal codec with auto-regressive entropy model~\cite{li2021deep} as described in Section~\ref{STVC}. The codec structure remains consistent in both intra- and inter-frame modes, wherein the spatio-temporal importance is guided by the reference feature $\hat{y}_\text{ref}$ during inter-frame compression. The codec initially unifies three types of frames and provides reference information for inter-frame coding without the additional modules.

Specifically, by using different reference features $\hat{y}_\text{ref}$, three types of frames can be preliminary unified into one framework. During the I-frame compression, as illustrated in the solid line of Figure~\ref{me1}, reference feature $\hat{y}_\text{ref}$ is not used. In P-frame compression, the previous decoded feature $\hat{y}_{i-1}$ is served as reference feature $\hat{y}_{\text{ref}}$, i.e. $\hat{y}_{\text{ref}}=\hat{y}_{i-1}$. For B-frame compression, the reference feature $\hat{y}_{\text{ref}}$ is synthesized by the previous decoded feature $\hat{y}_{i-1}$ and the following decoded feature $\hat{y}_{i+1}$, formulated as:
\begin{equation}
\hat{y}_{\text{ref}}=O \cdot \hat{y}_{i-1}+(1-O) \cdot \hat{y}_{i+1},
\end{equation}
where $O\in[0,1]$ represents the occlusion coefficient between bi-directional features $\hat{y}_{i-1}$ and $\hat{y}_{i+1}$. 

Relying solely on the above conditional coding makes it difficult to fully utilize inter-frame information to reduce bit-rate and reconstruct high-quality frames. Hence, three types of frames are reconstructed using conditional generation. The fusion feature $\hat{y}_i$ is used as the condition to perform Markovian dynamics $p_\theta(y_i^{0:T})$ on the initial state $y_i^T$ between a sequence of denoising steps $t=T\rightarrow 1 (T=30)$, formulated as:
\begin{equation}
p_\theta(y_i^{0:T} {\mid} \hat{y}_i)= p(y_i^T) \prod_{t=1}^T p_\theta(y_i^{t-1} {\mid} y_i^t, \hat{y}_i),
\end{equation}
where $p(y_i^T)=\mathcal{N}(y_i^T;0,1)$ represents the initial state of the transition for intra-frame compression. And $p(y_i^T)=\mathcal{N}(\hat{y}_{\text{ref}}^{T^{\prime}}; \mu_\theta(\hat{y}_{\text{ref}}^{T^{\prime}}, T^{\prime}), \sigma_\theta^2(\hat{y}_{\text{ref}}^{T^{\prime}}, T^{\prime}))$ denotes to conduct $\hat{y}_{\text{ref}}$ for DDIM inversion to $\hat{y}_{\text{ref}}^{T^{\prime}}$ as the initial state in inter-frame compression. $T^{\prime}=\frac{1}{2}T$ means the inverse noise level. $p_\theta(y_i^{t-1} {\mid} y_i^t)$ denotes the transition kernel through pre-trained U-Net~\cite{ronneberger2015u} in LDM~\cite{rombach2022ldm}, formulated as:
\begin{equation}
p_\theta(y_i^{t-1} {\mid} y_i^t)=\mathcal{N}(y_i^{t-1} ; \mu_\theta(y_i^t, t, \hat{y}_i), \sigma_\theta^2(y_i^t, t, \hat{y}_i)),
\label{transition}
\end{equation}
where the fusion feature $\hat{y}_i$ and the initial state $\hat{y}^{T^{\prime}}_{\text{ref}}$ share similar spatial and semantic structure, differing only in temporal motion dynamics. During the approximation for the mean $\mu_\theta$ and variance $\sigma_\theta^2$ of the target feature $y_i$, $y_i^t$ is aligned to $y_i^0$ conditioned on $\hat{y}_i$, achieving implicit inter-frame feature alignment. During inter-frame compression, this strategy incorporates inter-frame correlations to reduce the bit-rate, achieving high-quality frame reconstruction without MEMC. The details are described in Section~\ref{IIFA}. At last, a $4\times$ up-sampling $\mathcal{D}(\cdot)$ of pre-trained LDM is performed on the maximum likelihood feature $y^0_i$ to obtain the output frame $\hat{x}_i$.

\subsection{Spatio-Temporal Variable-rate Codec}
\label{STVC}

\begin{figure}[!t]
  \centering
  \includegraphics[width=\linewidth]{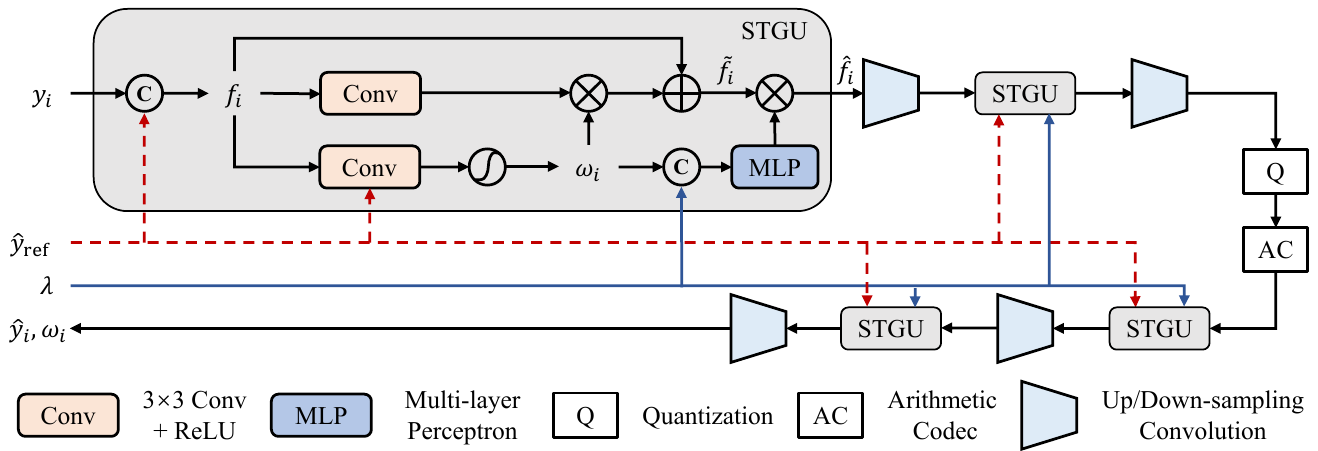}
  \caption{Overview of the Spatio-Temporal Variable-rate Codec (STVC). The target feature $y_i$ is scaled and transmitted as the fusion feature $\hat{y}_i$ with spatio-temporal attention on the reference feature $\hat{y}_\text{ref}$. The outputs also include the spatio-temporal importance-guided mask $\omega_i$.}
  \label{me2}
\end{figure}

As described in Section~\ref{VC3} and Figure~\ref{me1}, we propose a Spatio-Temporal Variable-rate Codec (STVC) to redefine the correlation of intra- and inter-frame in a unified network. Specifically, four Spatio-Temporal Guidance Units (STGU) are conducted to use the significance of reference features across each frame type. For intra-frame compression, only the self-spatial importance of the target feature $y_i$ is adopted for feature compression, as shown in Figure~\ref{me2}. To learn long-term dynamics from historical observations in inter-frame compression, the reference feature $\hat{y}_\text{ref}$ is deployed to perform spatio-temporal attention in STGU. Initially, the spatio-temporal importance-guided mask $\omega_i$ is conditioned on $\hat{y}_\text{ref}$, formulated as:
\begin{equation}
\omega_i = \sigma(\operatorname{Conv}(\operatorname{Concat}(f_i,\hat{y}_\text{ref}))),
\label{eq5}
\end{equation}
where $f_i$ is the concatenation of $\hat{y}_\text{ref}$ and $y_i$. $\operatorname{Concat}(\cdot)$ denotes the concatenation function. $\operatorname{Conv}(\cdot)$ denotes a 3$\times$3 convolution with ReLU. $\sigma(\cdot)$ means the sigmoid function. The spatial correlations and temporal dynamics of the target feature $y_i$ are selected by the mask $\omega_i$ as $\tilde{f}_i$:
\begin{equation}
\tilde{f}_i = f_i + \omega_i \cdot \operatorname{Conv}(f_i).
\label{eq6}
\end{equation}
The above Equation \ref{eq5} and Equation \ref{eq6} constitute the spatio-temporal attention mechanism, which preliminary introduces reference features in the form of conditional coding. Besides, the mask $\omega_i$ is employed to scale the feature $\tilde{f}_i$ with the variable-rate coefficient $\lambda$, formulated as:
\begin{equation}
\hat{f}_i = \operatorname{MLP}(\operatorname{Concat}(\omega_i,\lambda)) \cdot \tilde{f}_i,
\end{equation}
where $\operatorname{MLP}(\cdot)$ means the multi-layer perceptron. $\hat{f}_i$ denotes the scaled feature that preserves spatio-temporal deformations for subsequent implicit inter-frame alignment.

\subsection{Implicit Inter-frame Feature Alignment}
\label{IIFA}

After obtaining decoded features, we use conditional diffusion to generate three types of frames. It can integrate intra- and inter-frame dependency, but inevitably cause the limitation of temporal consistency. Therefore, as illustrated in Figure~\ref{me1} and Algorithm~\ref{algo1}, DDIM inversion is implemented to introduce inter-frame dependency for Implicit Inter-frame Feature Alignment (IIFA) without MEMC.

In detail, the DDIM inversion in inter-frame compression on the reference feature $\hat{y}_{\text{ref}}$ is guided by the spatio-temporal mask $\omega_i$, formulated as:
\begin{equation}
\hat{y}_{\text{ref}}^t=\sqrt{\alpha^t} \frac{\hat{y}_{\text{ref}}^{t-1}-\sqrt{1-\alpha^{t-1}} \omega_i\varepsilon_\theta}{\sqrt{\alpha^{t-1}}}+\sqrt{1-\alpha^t} \omega_i\varepsilon_\theta,
\label{ddim}
\end{equation}
where $\hat{y}^t_\text{ref}$ represents the noised feature that has undergone one DDIM inversion based on $\hat{y}^{t-1}_\text{ref}$, with the initial state $\hat{y}^0_\text{ref}=\hat{y}_{\text{ref}}$. $\alpha^{t-1}$ and $\alpha^t$ denote the weight coefficients during the inversion kernel. While a sequence of inverse steps (set as $t=1\rightarrow T^{\prime}$), IIFA retains similar low-frequency contextual structures and adds noise to high-frequency motion dynamics between the reference feature $\hat{y}_{\text{ref}}$ and the target feature $y_i$. Then, the initial state $y_i^T\backsim\mathcal{N}(0,1)$ in I-frame compression for denoising transition is replaced by the final noised feature $\hat{y}^{T^{\prime}}_\text{ref}$ in the inter-frame compression. 

This strategy allows step-by-step implicit temporal variations compensation from $y^T_i$ to $y^0_i$ using the denoising kernel (Equation~\ref{transition}) conditioned on the fusion feature $\hat{y}_i$, which includes spatial correlations from $y_i$ and temporal dynamics from $\hat{y}_\text{ref}$. Besides, the inverse steps $T^{\prime}$ is related to the motion complexity between $\hat{y}_{\text{ref}}$ and $y_i$. Therefore, we set $T^{\prime}=\frac{1}{2} T$ to improve the robustness of the inter-frame compression model. More analysis and discussion on implicit inter-frame alignment are described in Appendix~\ref{ad}.

% Consequently, by employing the masked DDIM inversion on the reference feature $\hat{y}_{\text{ref}}$ to obtain the initial state $y_i^T$ and utilizing $\hat{y}_i$ as the controllable condition of transition kernel, I$^2$VC adeptly integrates various inter-frame correlations into a unified framework for video compression. 

 \begin{figure}[!t]  
 \centering
 \begin{minipage}[b]{0.325\linewidth} 
   \centering
   {\includegraphics[width=\linewidth]{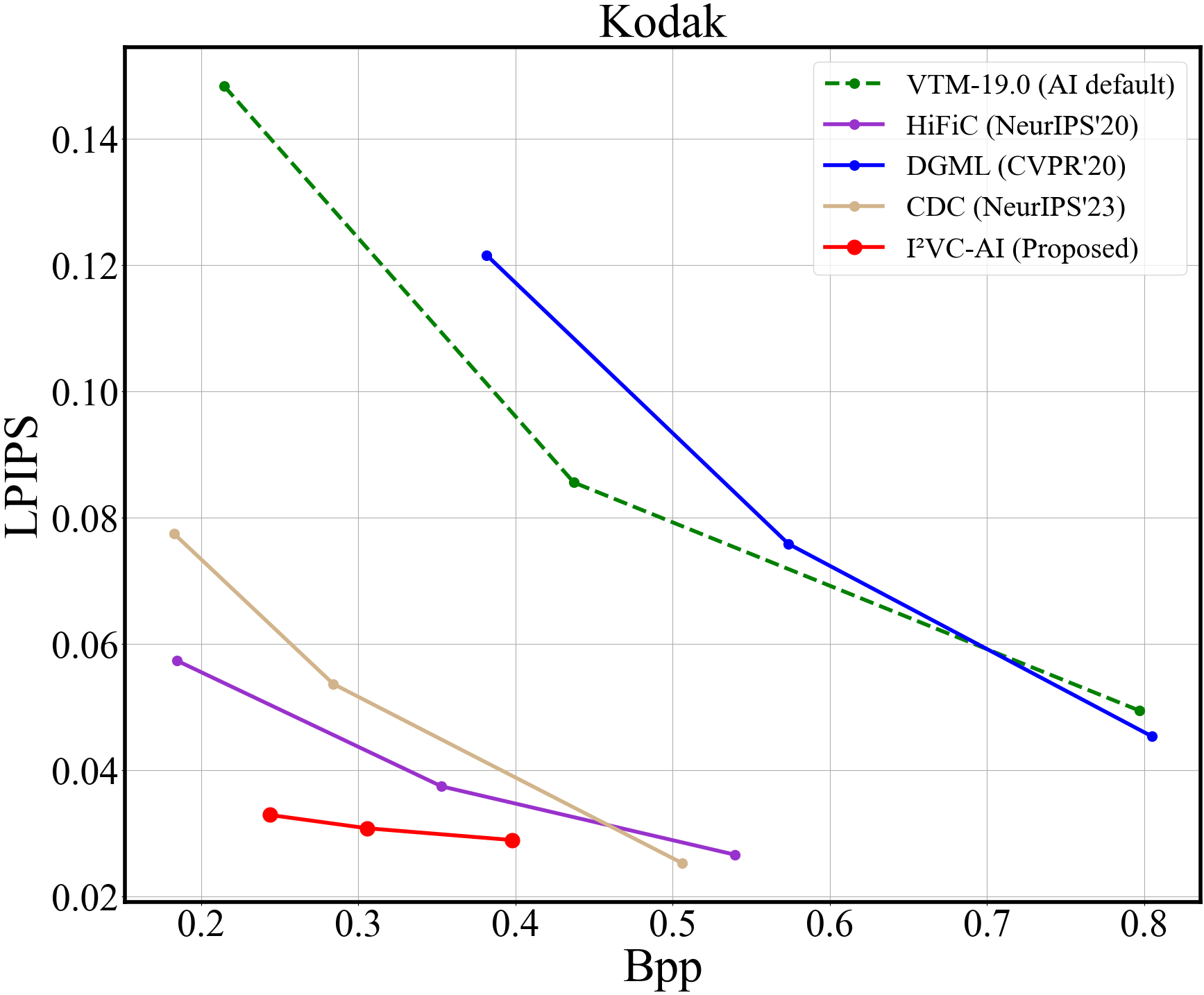}}
 \end{minipage}
 \hfill
 \begin{minipage}[b]{0.325\linewidth}
   \centering
   {\includegraphics[width=\linewidth]{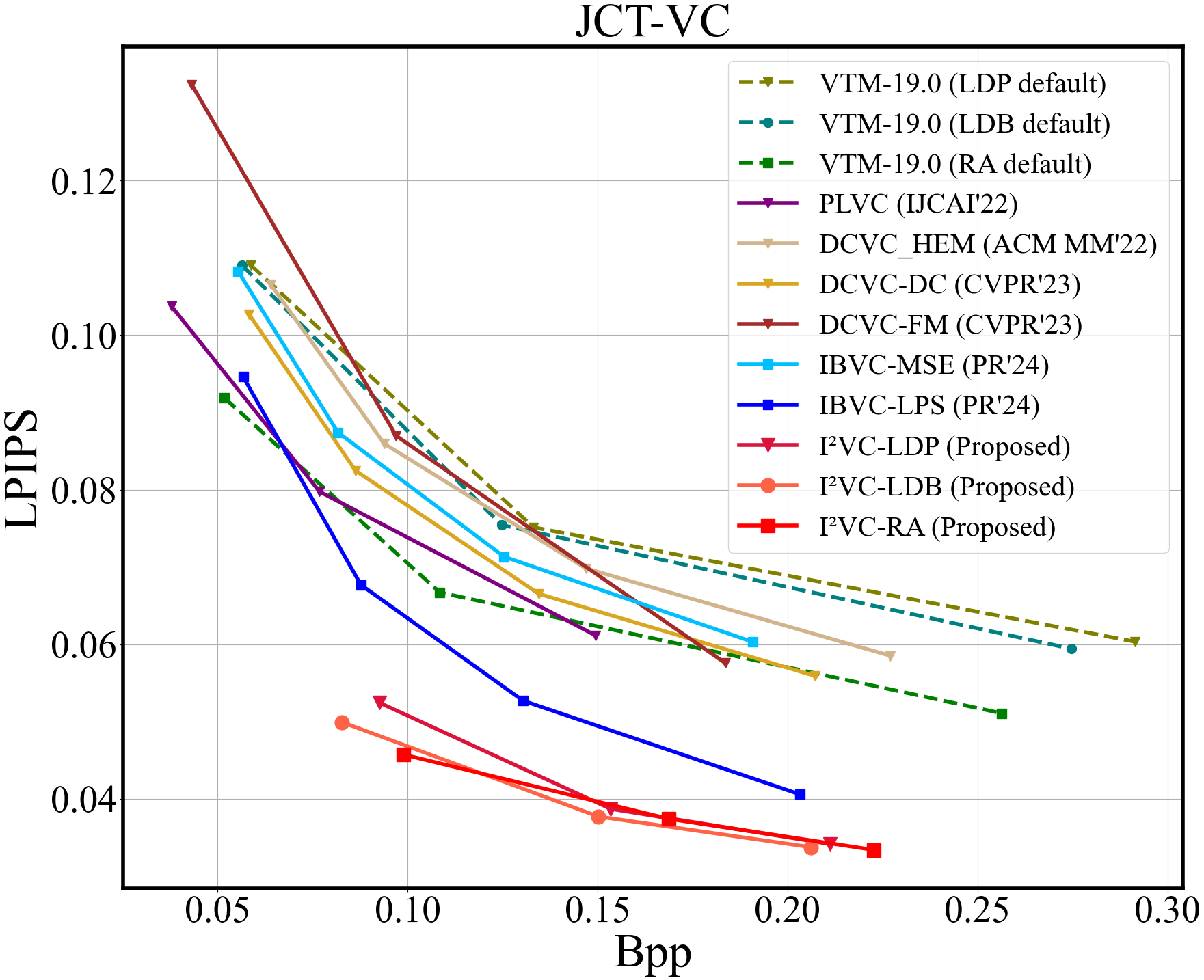}}
 \end{minipage}
 \hfill
 \begin{minipage}[b]{0.325\linewidth}
   \centering
   {\includegraphics[width=\linewidth]{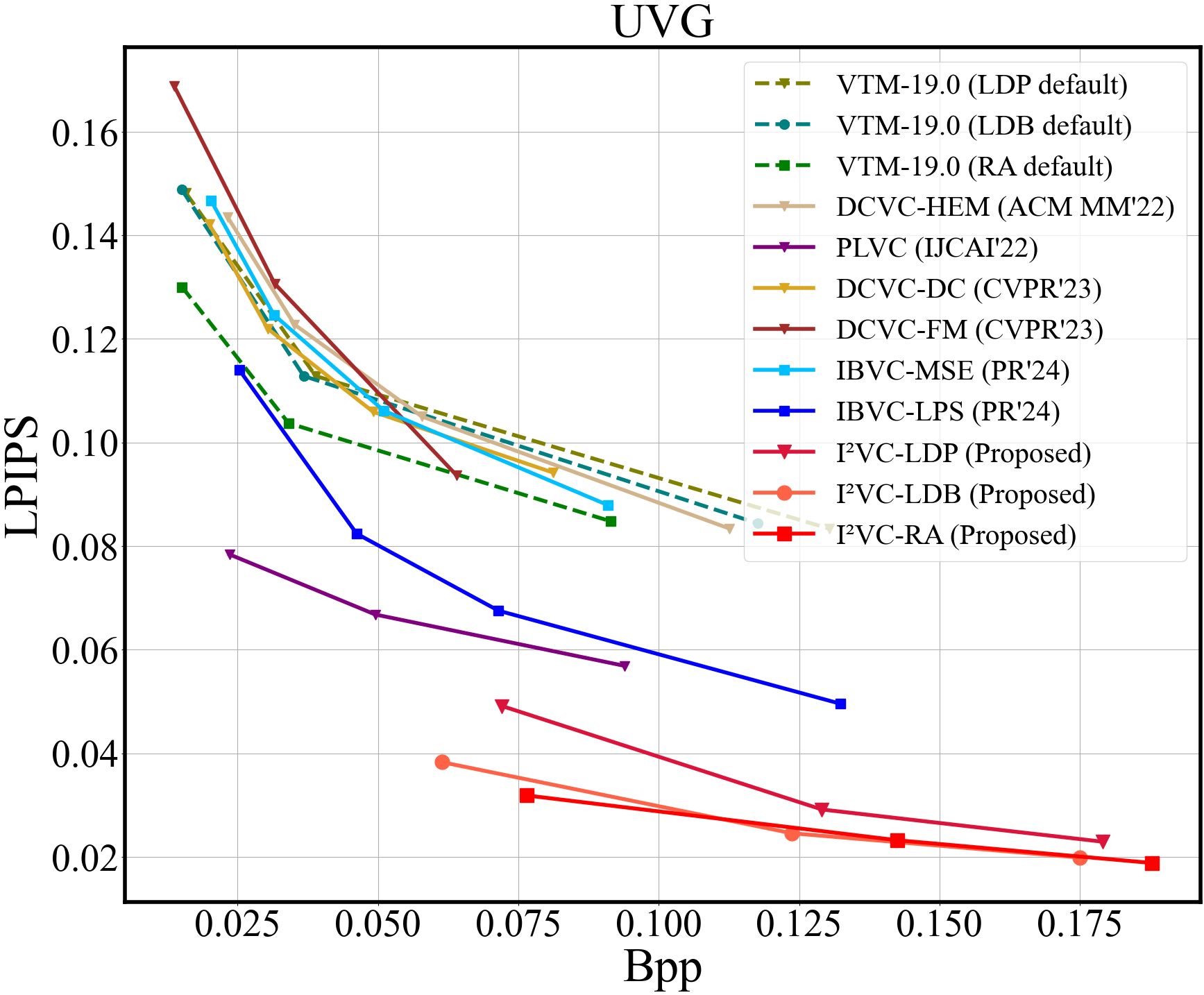}}
 \end{minipage}
   \caption{Rate-perception comparison of different video compression methods with three configurations on Kodak, JCT-VC, and UVG datasets in terms of LPIPS$\downarrow$.}
  \label{ex1}
\end{figure}

\subsection{Objective Function}

We use unified object functions $\mathcal{L}$ to ensure Rate-Distortion (R-D) and Rate-Perception (R-P) trade-off across different compression configurations, formulated as:
\begin{equation}
\mathcal{L} = \mathcal{R}(\hat{y}_i) + \lambda \cdot (\mathcal{D} (x_i, \hat{x}_i) + \beta \cdot \mathcal{P} (x_i, \hat{x}_i)),
\label{eq_opt}
\end{equation}
where $\lambda$ and $\beta$ denote the scale factors among bit-rate $\mathcal{R}$, distortion $\mathcal{D}$ and perception $\mathcal{P}$, respectively. Thanks to the proposed variable-rate codec, only one model is trained with $\lambda\in[8,512]$ for Mean Squared Errors (MSE) distortion and $\beta=0.05$ for Learned Perceptual Image Patch Similarity (LPIPS)~\cite{zhang2018unreasonable}. The specific training strategy is described in Appendix~\ref{ts}.

\section{Experiments}

\subsection{Settings}
\label{sec4.1}

We adopt the same training dataset Vimeo-90K Septuplet~\cite{xue2019video} as the previous method~\cite{li2021deep} for a fair comparison. The Kodak~\cite{kodak}, JCT-VC (HEVC Class B, C and D)~\cite{bossen2013common} and UVG ~\cite{mercat2020uvg} datasets are utilized to evaluate the performance of I$^2$VC. We make a comparison with traditional codec VTM-19.0~\cite{vtm} and state-of-the-art learned codecs across AI, LDP, LDB and RA configurations. Specifically, the learned I-frame compression method DGML~\cite{cheng2020learned}, learned generative I-frame compression method HiFiC~\cite{mentzer2020high} and CDC~\cite{yang2023cdc}, learned P-frame compression method DCVC-HEM~\cite{li2022hybrid}, DCVC-DC~\cite{li2023dcvcdc}, and DCVC-FM~\cite{li2024dcvcfm}, learned generative P-frame compression method PLVC~\cite{yang2022perceptual}, learned B-frame compression method IBVC~\cite{xu2024ibvc} and its LPIPS model IBVC-LPS~\cite{xu2024ibvc} are considered for comparisons. We test 32 frames of each video sequence for inter-frame compression. The GoP size is set as 32. Precisely, there are 32 I-frame in I$^2$VC-AI; 1 I-frame, 31 P-frame in I$^2$VC-LDP; 1 I-frame and 6 P-frame and 25 B-frame in I$^2$VC-LDB; 2 I-frame and 30 B-frame in I$^2$VC-RA. All experiments are performed on 4 NVIDIA GeForce RTX 3090 GPUs with Intel(R) Xeon(R) Gold 6248R CPUs. 

\subsection{Evaluation against State-of-the-art Methods}

\subsubsection{Quantitative Evaluations} 

\begin{figure}[!t]  
 \centering
 \begin{minipage}[b]{0.48\linewidth} 
   \centering
   {\includegraphics[width=\linewidth]{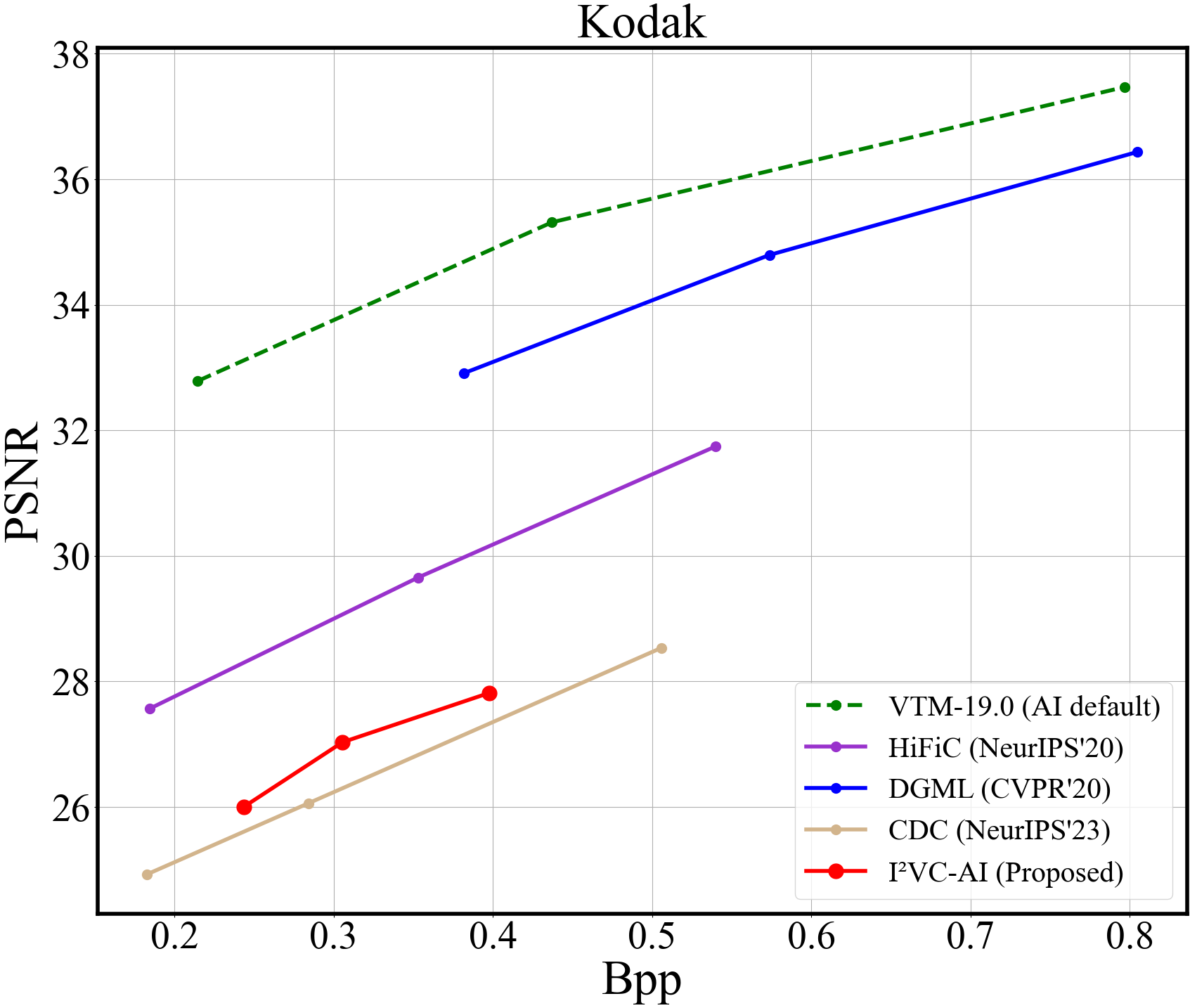}}
 \end{minipage}
 \hfill
 \begin{minipage}[b]{0.49\linewidth}
   \centering
   {\includegraphics[width=\linewidth]{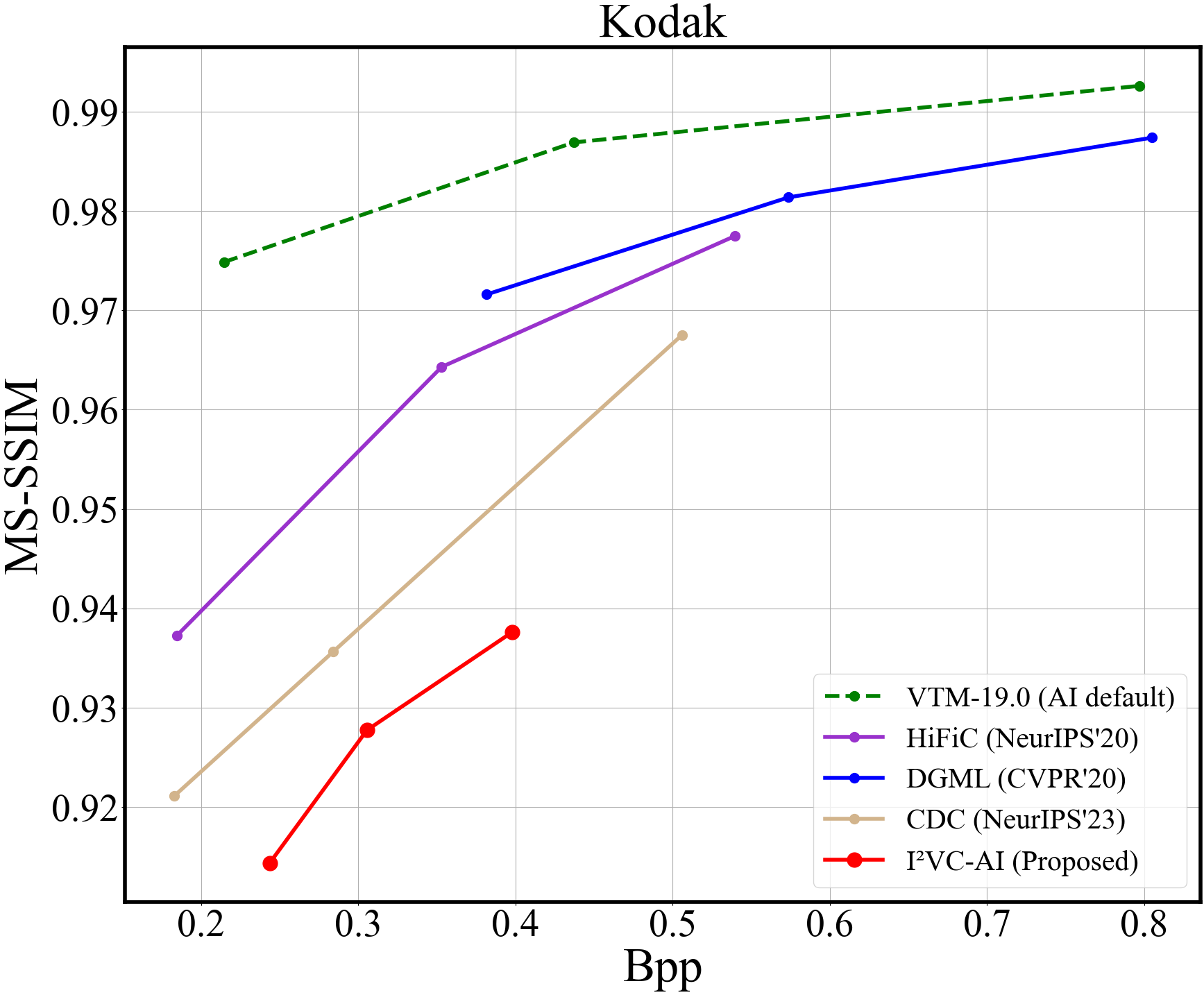}}
 \end{minipage}
   \caption{Rate-distortion comparison of different intra-frame compression methods on Kodak dataset in terms of PSNR$\uparrow$ and MS-SSIM$\uparrow$.}
  \label{psnrAndSSIM}
\end{figure}

It is noted in Figure \ref{ex1} that I$^2$VC achieves state-of-the-art R-P performance on test datasets compared to VTM-19.0~\cite{vtm} and deep generative compression methods~\cite{mentzer2020high,yang2022perceptual}. Especially, I$^2$VC reaches an average of 58.4\% perceptual improvements than VTM-19.0~\cite{vtm} with a similar bit-rate across different configurations. Besides, we compare the R-D performance of I$^2$VC-AI and other intra-frame compression methods on Kodak dataset. As displayed in Figure \ref{psnrAndSSIM}, I$^2$VC-AI can reach comparable MS-SSIM~\cite{wang2003msssim} and even better PSNR than the training required diffusion-based CDC~\cite{yang2023cdc}. Considering inter-frame compression, the pre-trained diffusion-based method~\cite{li2024extreme} achieves a reconstruction quality of 24.47dB using 0.06 Bits per pixel (Bpp) on UVG dataset, while I$^2$VC achieves an average reconstruction quality of 31.03dB across three configurations with the same bit-rate. It demonstrates that I$^2$VC maintains the better fidelity of reconstructed videos while leveraging pre-trained models. The experimental results indicate that I$^2$VC can achieve generalized perceptual reconstruction with acceptable fidelity across different configurations with a unified framework. Furthermore, we display Fréchet Inception Distance (FID)~\cite{heusel2017fid} performance in Figure~\ref{ap1} of Appendix~\ref{keguan}.  

\subsubsection{Qualitative Evaluation}

\begin{figure}[!t]
  \centering
  \includegraphics[width=\linewidth]{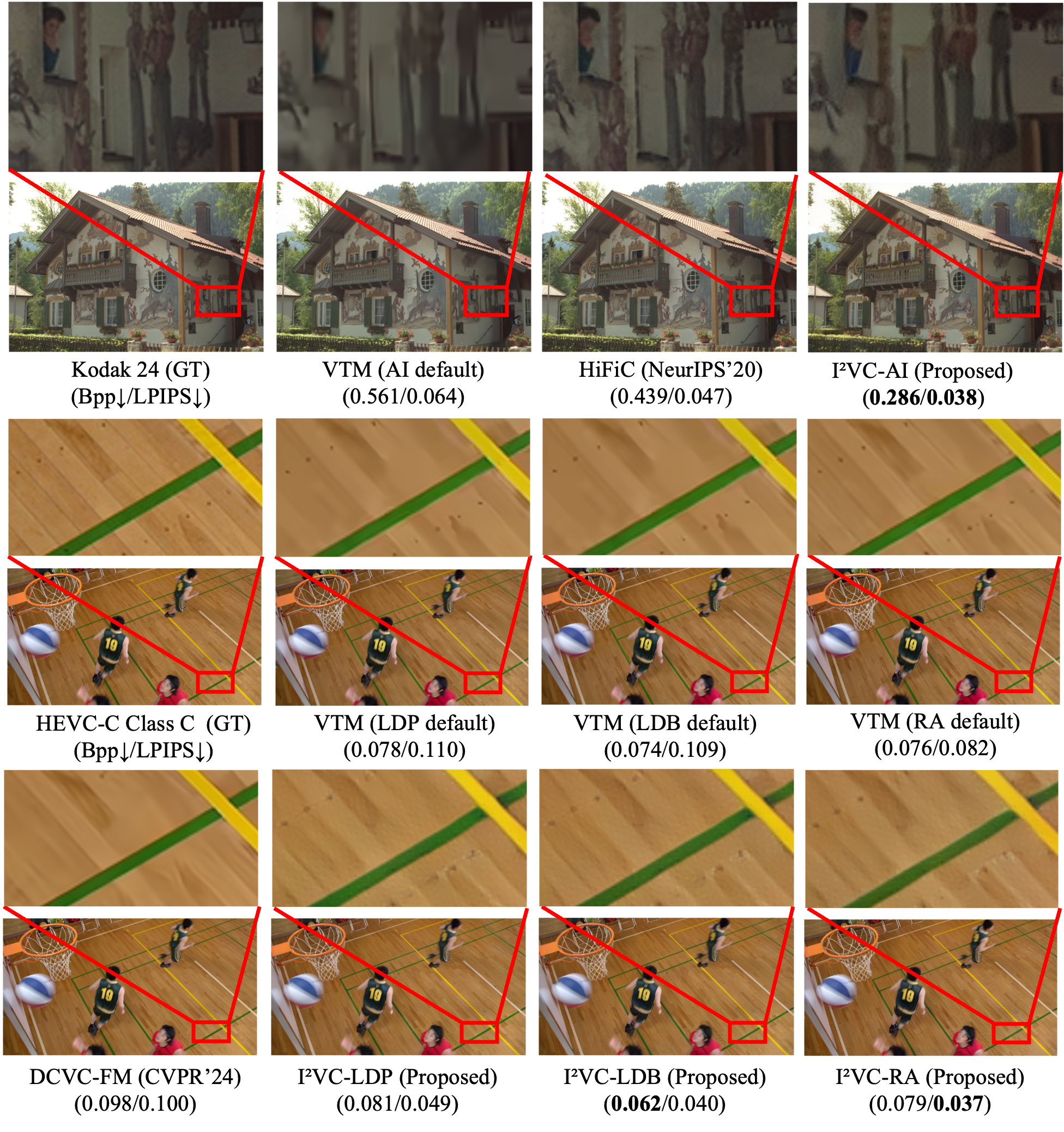}
  \caption{Qualitative comparison of different video compression configurations on Kodak and HEVC Class C datasets.}
  \label{example1}
\end{figure}

We provide a qualitative comparison of Kodak and HEVC Class C datasets between I$^2$VC and other methods \cite{vtm,li2024dcvcfm,mentzer2020high,yang2023cdc} as depicted in Figure \ref{example1}. For AI configuration on Kodak dataset, I$^2$VC exhibits fewer artifacts and achieves similar visual quality with a lower bit-rate compared to VTM-19.0 (AI default)~\cite{vtm}. For LD and RA configuration on Kodak dataset, I$^2$VC achieves richer visually pleasing textures than VTM-19.0~\cite{vtm} with similar bit-rate and DCVC-FM~\cite{li2024dcvcfm} with 1.3× the bit-rate of ours. It can be observed that VTM-19.0~\cite{vtm} loses some texture details of floor gaps in the second example. In contrast, our model achieves sharp boundaries and realistic textures without excessive smoothing and motion artifacts. It verifies the temporal consistency of I$^2$VC in inter-frame coding. We provide more visual examples in Figure~\ref{ap2} of Appendix \ref{zhuguan}.

\subsection{Ablation Study}

\subsubsection{Ablation of the Unified Framework for Inter-frame Video Compression}
\label{seab1}
\begin{figure}[!t] 
    \centering
    \begin{subfigure}[b]{0.3\textwidth}
        \includegraphics[width=\textwidth]{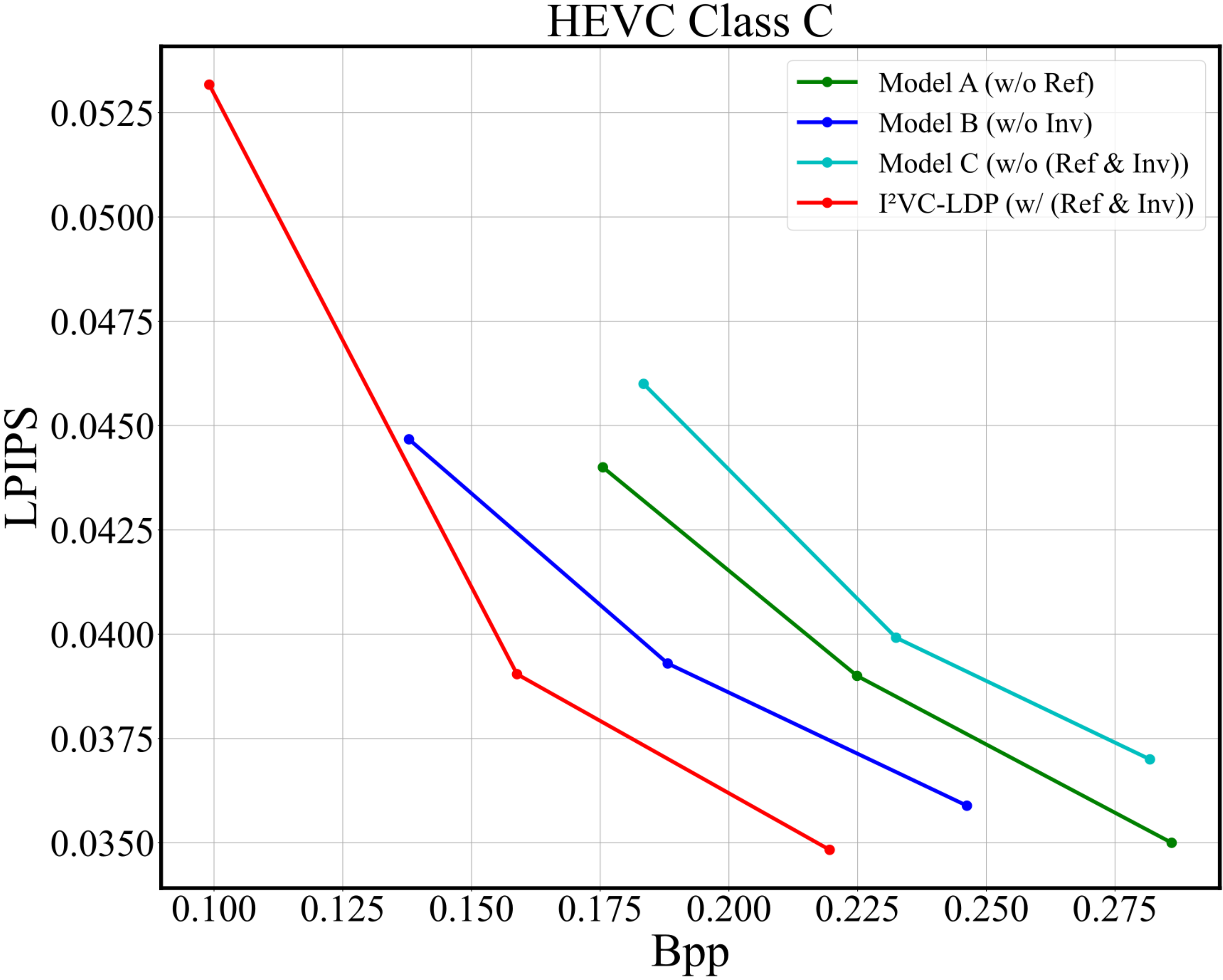}
        \caption{Ablation of the proposed inter-frame framework.}
    \end{subfigure}
    \hfill 
    \begin{subfigure}[b]{0.3\textwidth}
        \includegraphics[width=\textwidth]{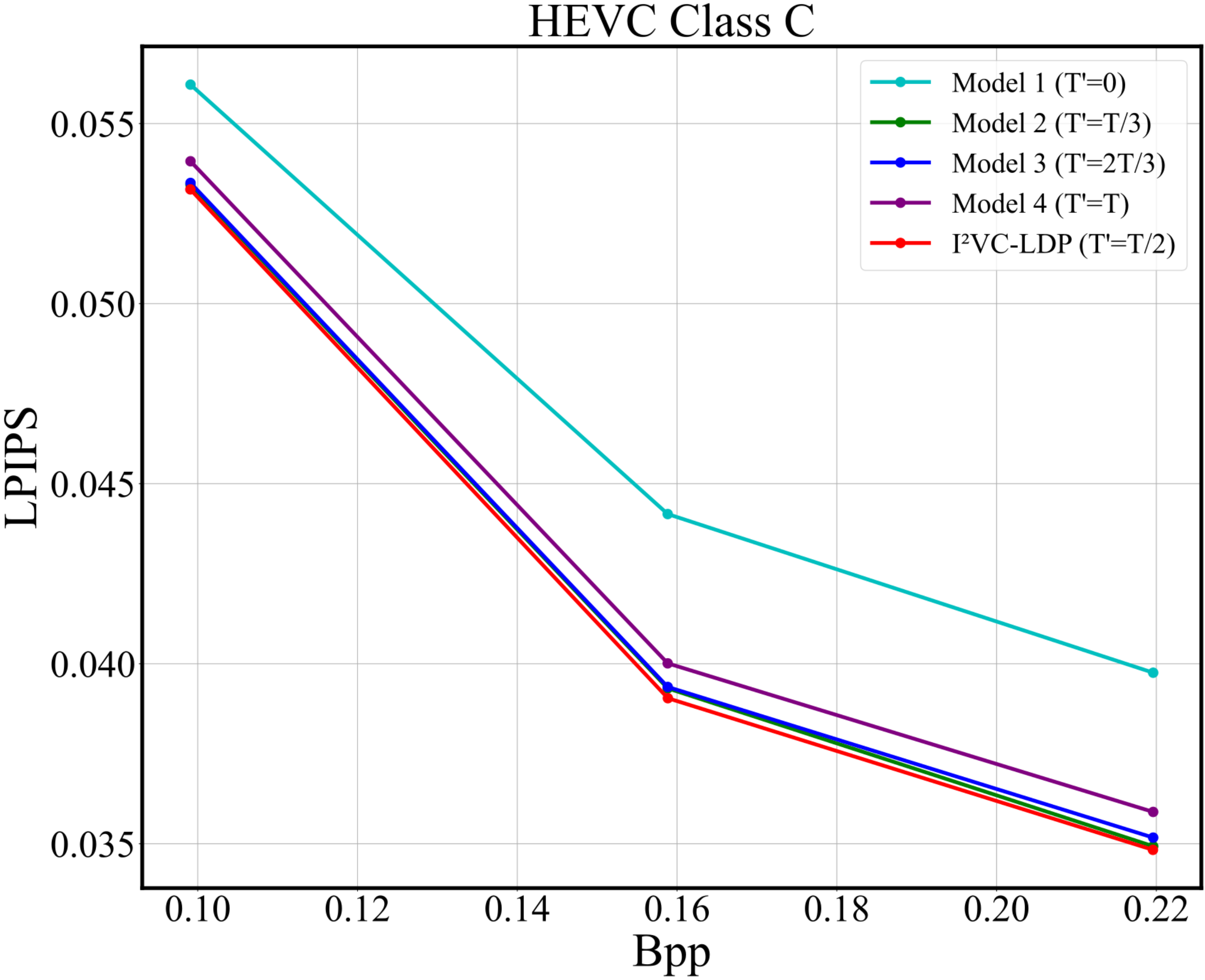}
        \caption{Ablation of the masked DDIM inversion steps.}
    \end{subfigure}
    \hfill 
    \begin{subfigure}[b]{0.3\textwidth}
        \includegraphics[width=\textwidth]{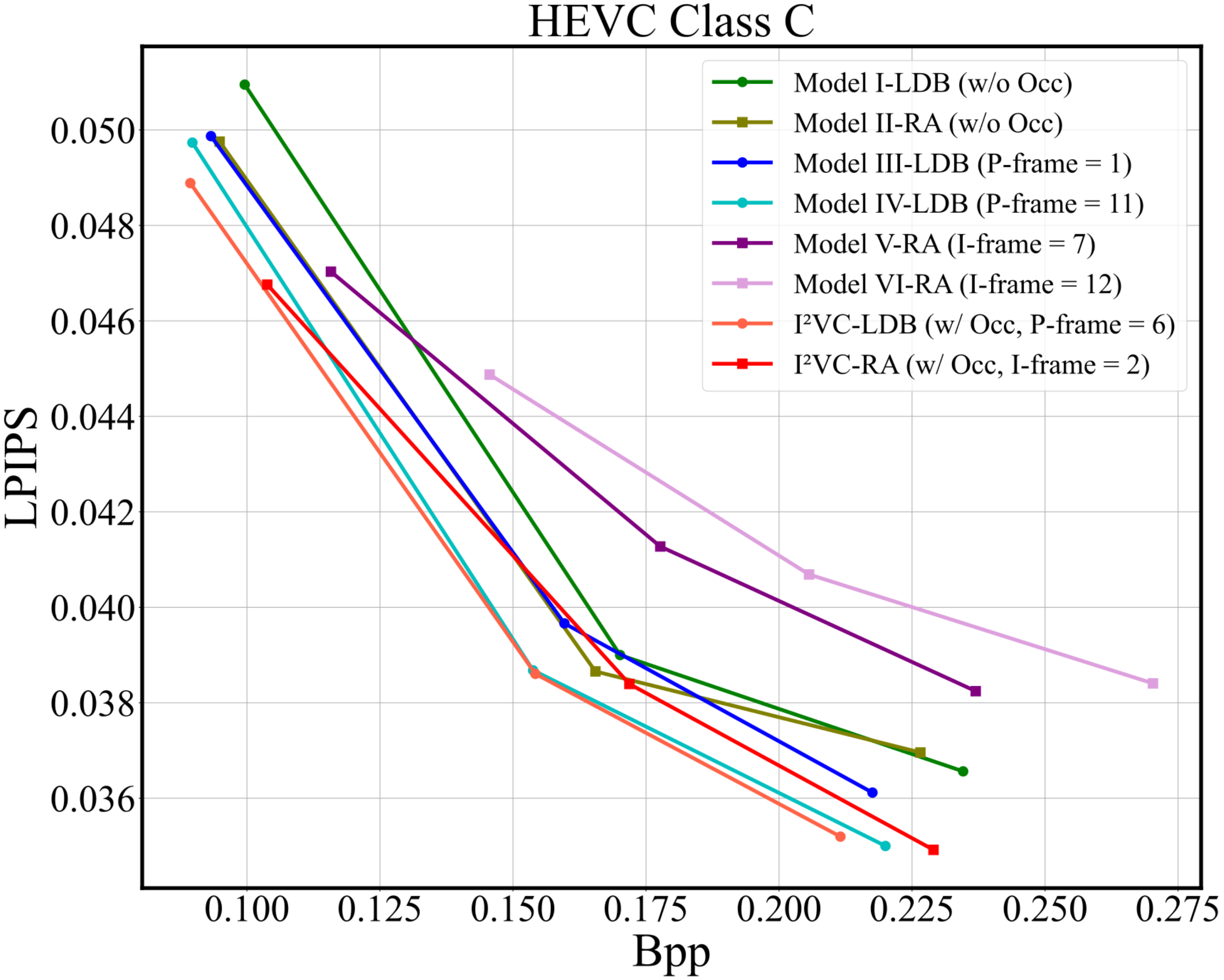}
        \caption{Ablation of the bi-directional inter-frame compression.}
    \end{subfigure}
    \caption{Ablations of the proposed inter-frame framework, the masked DDIM inversion steps, and the bi-directional inter-frame compression on HEVC Class C dataset.} 
    \label{ab1}
\end{figure}

To verify the effectiveness of I$^2$VC, we conduct three ablations on HEVC Class C dataset regarding the LDP configuration. Model A (w/o Ref) is designed to investigate the impact of STVC which utilizes reference features for unification. Concretely, $\hat{y}_{\text{ref}}$ is not used as a prior condition in Equation \ref{concat}, the inter-frame compression is replaced by intra-frame compression, as $\hat{y}_i=D(\left\lfloor E(y_i)\right\rceil)$. Model B (w/o Inv) is deployed to verify the effectiveness of IIFA on the reference feature $\hat{y}_{\text{ref}}$. Random noise $\varepsilon \sim N(0,1)$ is used as initial state for inter-frame compression. Additionally, we conduct Model C (w/o Ref \& Inv) without the reference feature $\hat{y}_{\text{ref}}$ for STVC and IIFA. As depicted in Figure \ref{ab1}(a), I$^2$VC-LDP saves bit-rate with the same reconstruction quality as the ablation models. And it indicates that the reference feature $\hat{y}_{\text{ref}}$ is vital to I$^2$VC to reduce the bit-rate and improve the perception quality.

% To explore prior inter-frame correlations for state transition, the masked DDIM inversion is adopted on the reference feature $\hat{y}_{\text{ref}}$ in place of random noise $\varepsilon \sim N(0,1)$ as the initial state.

\subsubsection{Ablation of the Implicit Inter-frame Feature Alignment}

\begin{figure}[!t]
  \centering
  \includegraphics[width=\linewidth]{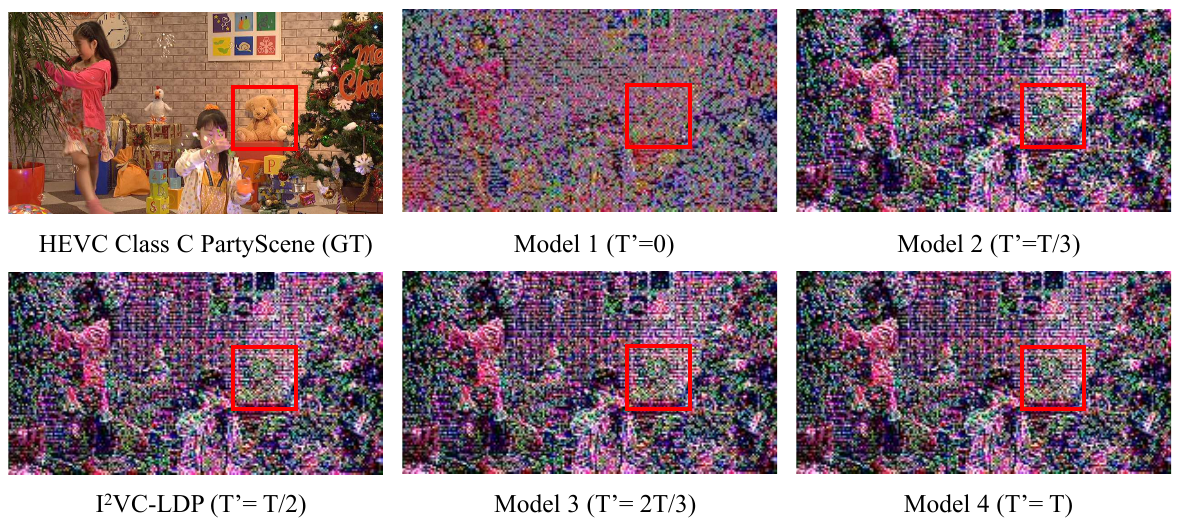}
  \caption{Visual example of $\hat{y}^{T^{\prime}}_{\text{ref}}$ about DDIM inversion from PartyScene in HEVC Class C dataset.}
  \label{Ablation_Qualitative}
\end{figure}

Model B (w/o Inv) in Section \ref{seab1} has confirmed that the DDIM inversion can introduce reference information to optimize the denoising process. Furthermore, Model 1 to 4 under the LDP configuration on HEVC Class C dataset with DDIM inversion steps of $0, \frac{1}{3} T, \frac{2}{3} T$, and $T$ are conducted to validate the impact of DDIM inversion for IIFA. As shown in Figure \ref{ab1}(b), I$^2$VC-LDP with $T^{\prime}=\frac{1}{2} T$ achieves the best R-P performance compared to Models 1 to 4. As depicted in Figure \ref{Ablation_Qualitative}, I$^2$VC-LDP ($T^{\prime}=\frac{1}{2}T$) has a clearer structural definition than Model 1 ($T^{\prime}=0$) and preserves more semantic contents than Model 4 ($T^{\prime}=T$). Therefore, the appropriate number of DDIM inversion steps on the reference feature $\hat{y}_{\text{ref}}$ enables to add noise on the temporal dynamics with structural maintenance, providing a reference for inter-frame alignment in subsequent transition. Besides, we conduct an ablation of direct frame reconstruction from the fusion feature $\hat{y}_i$ without diffusion denoising, but the reconstruction performance is unacceptable and therefore not presented. Therefore, it demonstrates that the proposed IIFA can fuse multi-frame information for video compression.

\subsubsection{Ablation of Bi-directional Inter-frame Compression}

To investigate the impact of the occlusion coefficient on bi-directional inter-frame compression, Model I-LDB (w/o Occ) and Model II-RA (w/o Occ) is set up on HEVC Class $C$ dataset. Both models exclude occlusion coefficients for direct fusion of bi-directional inter-frame reference features. As shown in Figure \ref{ab1}(c), I$^2$VC-LDB and I$^2$VC-RA achieve the same perceptual quality while bit-rate saving compared to Model I and Model II. It demonstrates the effectiveness of the occlusion coefficient in the fusion of bi-directional inter-frame reference information.

To investigate the influence of the GoP structures on the bi-directional compression, we vary the number of P-frame in the LDB configuration and I-frame in the RA configuration. I$^2$VC-LDB (P-frame=6) has GOP length of 32 with 6 P-frame. Model III-LDB (P-frame=1) has a GoP length of 32 with 1 P-frame. Model IV-LDB (P-frame=11) has a GoP length of 32 with 11 P-frame. As depicted in Figure \ref{ab1}(c), I$^2$VC-LDB achieves the optimal reconstruction quality, especially at a lower bit-rate. The experimental results demonstrate that an appropriate GoP structure can regulate the distance between the reference frame and the target frame, ensuring a trade-off between bit-rate and reconstruction quality. The GoP length of I$^2$VC-RA (I-frame=2), Model V-RA (I-frame=7) and Model VI-RA (I-frame=12) is 32, and the number of I-frame is 2, 7 and 12, respectively. As shown in Figure \ref{ab1}(c), I$^2$VC-RA achieves the best reconstruction results, saving about 33\% bit-rate compared to Model V with the same reconstruction quality. It demonstrates that I$^2$VC-RA can sufficiently leverage bi-directional inter-frame correlations to save bit-rate.

\section{Conclusion and Limitation}
\label{sec5}

Current video compression methods require training distinct frameworks for different configurations with three types of frames, resulting in model redundancy and weak generalization. Thus, we propose a unified framework for Intra- \& Inter-frame Video Compression (I$^2$VC), including a single spatio-temporal variable-rate codec and an implicit inter-frame alignment. Initially, the unified codec inter-frame dependency into conditional coding referred to decoded reference features, preliminary integrating intra- and inter-frame correlations to one framework. Subsequently, to resolve the absence of motion information, reference features are applied for DDIM inversion as the initial state of diffusion denoising instead of random noise. It allows for selective denoising of motion-rich regions based on decoded features, facilitating implicit inter-frame alignment without MEMC. The experimental results across three configurations demonstrate that I$^2$VC achieves an average of 58.4\% perceptual improvements over VTM-19.0 with the same bit-rate and exhibits superior R-P performance compared to other learned video compression methods. However, I$^2$VC primarily relies on pre-trained LDM constrained by training complexity and device memory. As a result, it exhibits certain limitations in terms of different metrics. Future research will aim to train the complete diffusion and inverse process to enhance reconstruction performance and generalization.

\newpage

\bibliographystyle{plain}
\bibliography{neurips_2024}

\newpage

\appendix

\section{Analysis and Motivation}
\label{pm}

\subsection{Learned Video Compression}

Current learned video compression methods~\cite{balle2018variational,lu2019dvc,yang2020Learning} leverage various inter-frame dependencies to achieve three different coding configurations (AI, LD and RA), catering to different scenarios. Specifically, the AI configuration is often used for video sequences with sparse inter-frame correlation and low frame rates. Each frame is an I-frame and independently compressed without reference from other frames. The VAE codec~\cite{balle2018variational} is used in I-frame compression for output frame $\hat{x}_i$, formulated as:
\begin{equation}
\hat{x}_i = {D}(\lfloor {E}(x_i)\rceil),
\label{eq3.1}
\end{equation}
where $E(\cdot)$ denotes the encoder to compress input frame $x_i$ as latent feature. $D(\cdot)$ denotes the decoder which restores latent feature back to $\hat{x}_i$. 

The LD configuration is used for video coding in low-latency scenarios. Three types of frames (I-frame, P-frame and B-frame) are included in this configuration. The RA configuration is used for random access within video streams, comprising I-frame and B-frame, which respectively utilize unidirectional and bi-directional inter-frame references. A hybrid compression framework incorporating motion codec and residual codec~\cite{lu2019dvc,yang2020Learning} is employed in P-frame and B-frame compression, the predicted frame $\tilde{x}_i$ and output frame $\hat{x}_i$ are respectively formulated as:
\begin{equation}
\begin{aligned}
& \tilde{x}_i=W(x_{i-1},\left[x_{i+1}\right], D_\text{m}(\left\lfloor E_\text{m}(m_{x_{i-1} \rightarrow x_i},\left[m_{x_{i+1} \rightarrow x_i}\right])\right\rceil)), \\
& \hat{x}_i=D_\text{r}(\left\lfloor E_\text{r}(x_i-\tilde{x}_i)\right\rceil)+\tilde{x}_i,
\end{aligned}
\label{eq3.2}
\end{equation}
where $E_\text{m}(\cdot)$ and $D_\text{m}(\cdot)$ represent the motion codec to transmit the motion information between reference frame ${x_{i-1},[x_{i+1}]}$ and input frame $x_i$ to obtain predicted frame $\tilde{x}_i$. ${E}_\text{r}(\cdot)$ and ${D}_\text{r}(\cdot)$ represent the residual codec to transmit the residual information between predicted frame $\tilde{x}_i$ and input frame $x_i$ to obtain output frame $\hat{x}_i$. $[\cdot]$ indicates the variable is optional.

Though there are similarities between  Equation~\ref{eq3.1} and Equation~\ref{eq3.2}, the inter-frame video compression framework includes an extra motion codec compared to the intra-frame framework. Furthermore, there are also significant differences within the inter-frame video compression frameworks. P-frame compression necessitates only unidirectional reference frames, but B-frame compression requires bi-directional reference frames, as shown in Equation~\ref{eq3.2}. The MEMC and motion codec in the B-frame coding framework are more complex, making these three frameworks incompatible.

\subsection{Video Diffusion Model}

Existing video restoration and generation methods~\cite{chen2023motion, voleti2022mcvd,danier2024ldmvfi, peng2023conditionvideo, hu2023videocontrolnet} implement DDIM~\cite{song2020DDIM} to achieve the fusion of inter-frame dependency. During the process of diffusion denoising, the maximization distribution $p_\theta(x_i^0)$ with variables $x_i^{1:T}$ of input frame $x_i$ is formulated as:
\begin{equation}
p_\theta(x_i^0)=\int p_\theta(x_i^{0:T}) d x_i^{1: T},
\end{equation}
where $p_\theta(x_i^{0:T})$ denotes the Markovian dynamics between a sequence of transitional steps $t=T\rightarrow1$, which can be specifically formulated as:
\begin{equation}
p_\theta(x_i^{0:T})=p(x_i^T) \prod_{t=1}^T p_\theta(x_i^{t-1} {\mid} x_i^t),
\label{eq3.3}
\end{equation}
where $p(x_i^T)=\mathcal{N}(x_i^T;0,1)$ represents the initial distribution with random Gaussian noise. $p_\theta(x_i^{t-1} {\mid} x_i^t)$ represents the state transition kernel using U-Net~\cite{ronneberger2015u}, formulated as:
\begin{equation}
p_\theta(x_i^{t-1} {\mid} x_i^t)=\mathcal{N}(x_i^{t-1} ; \mu_\theta(x_i^t, t), \sigma_\theta^2(x_i^t, t)),
\end{equation}
where $\mu_\theta(x_i^t, t)$ and $\sigma_\theta^2(x_i^t, t)$ denote the mean and variance of the transition kernel. To better estimate the distribution $p_\theta(x_i^{t-1} {\mid} x_i^t)$, the method~\cite{voleti2022mcvd} introduces reference frames as the controllable condition, formulated as:
\begin{equation}
p_\theta(x_i^{t-1} {\mid} x_i^t)=\mathcal{N}(x_i^{t-1} ; \mu_\theta(x_i^t, t, \tau_\theta), \sigma_\theta^2(x_i^t, t, \tau_\theta)),
\label{original_denoise}
\end{equation}
where $\tau_\theta$ denotes the controllable condition, which can be expressed as forward reference frame $x_{i-1}$ or bi-directional reference frame $\left\{x_{i-1}, x_{i+1}\right\}$ for video restoration in~\cite{voleti2022mcvd}. However, using only reference frames as controllable conditions introduces higher uncertainty, leading to the essential training with diffusion process $q_\theta(x_i^{t} {\mid} x_i^{t-1})$. Other methods use motion vector~\cite{chen2023motion}, optical flow~\cite{hu2023videocontrolnet} and depth image~\cite{hu2023videocontrolnet}, etc. combined with multi-modal information as the controllable condition $\tau_\theta$, but resulting in unrealistic video reconstruction with weak consistency.

\section{Related Work}
\label{relate}
Currently, some methods commonly utilize Denoising Diffusion Probabilistic Models (DDPM)~\cite{ho2020DDPM} and DDIM~\cite{song2020DDIM} to model the inter-frame correlations between video frames. Yang \emph{et al.}~\cite{yang2023diffusion} directly utilize DDPM to predict future video frames. Harvey \emph{et al.}~\cite{harvey2022flexible} propose a diffusion model to generate long-range video conditioned on any sampled arbitrary subset of original video frames. Voleti \emph{et al.}~\cite{voleti2022mcvd} introduce a general video synthesis probabilistic diffusion conditioned on past and/or future frames. Yang \emph{et al.}~\cite{yang2023vdm} present a local-global video diffusion model using 3D convolution to capture multi-perception conditions for video synthesis. Danier \emph{et al.}~\cite{danier2024ldmvfi} propose a video frame interpolation framework by latent diffusion models. Mei \emph{et al.}~\cite{mei2023vidm} devise a video implicit diffusion model to generate motion using the latent map of the first and latest frames. Chai \emph{et al.}~\cite{chai2023stablevideo} utilize a pre-trained propagator in diffusion for video editing. Yu \emph{et al.}~\cite{yu2023video} efficiently transform video to low-dimensional projected latent space for training the diffusion model within limited resources. 

Some methods leverage multi-modal conditions to incorporate inter-frame dependency relationships. Ho \emph{et al.}~\cite{ho2022imagen, Ho2022VDM} first propose a diffusion-based text-to-video model and develop a large text-conditioned video generation task. Chen \emph{et al.}~\cite{chen2023motion} use intended content and dynamics from users as the diffusion condition to synthesize video. The above methods are limited by high training complexity. Therefore, some methods~\cite{wu2023tune,khachatryan2023text2video,lu2023tf} extend the pre-trained diffusion models~\cite{rombach2022ldm,zhang2023controlnet} by incorporating inter-frame temporal information using cross-attention. Furthermore, Esser \emph{et al.}~\cite{esser2023structure} control structure by input video and content by input text based on pre-trained image diffusion. Hu \emph{et al.}~\cite{hu2023videocontrolnet} use optical-flow as a condition of ControlNet for video-to-video translation. Hu \emph{et al.}~\cite{hu2023lamd} introduce a motion generator by latent diffusion model for video generation. Peng \emph{et al.}~\cite{peng2023conditionvideo} disentangle the inter-frame dynamics into scenery motion components, such as pose, depth, and segmentation which can be the condition for pre-trained ControlNet. Moreover, some methods~\cite{ceylan2023pix2video,liao2023lovecon,qi2023fatezero} introduce DDIM inversion to ensure the inter-frame consistency of invariant content while modifying only the edited content in video editing. However, the multi-modal conditions and high complexity of cross-attention pose challenges for the aforementioned methods in effectively recovering the original video in video compression tasks.

\begin{algorithm}[!t]
\caption{A unified framework for intra- \& inter-frame video compression}\label{algo1}
\begin{algorithmic}[1]
\renewcommand{\algorithmicrequire}{ \textbf{Input:}}   
\renewcommand{\algorithmicensure}{ \textbf{Output:}}   
\Require $x_i,\left[\hat{y}_{i-1}\right],\left[\hat{y}_{i+1}\right]$~~ {// Input frame and reference features from feature buffer.}
\Ensure $\hat{x}_i, \hat{y}_i$~~ {// Output frame and fusion feature for feature buffer.}
\State $y_i=\mathcal{E}(x_i)$~~ {// Feature domain.}
\State // Frame types judgment for reference feature $\hat{y}_{\text{ref}}$ and the initial state $y_i^T$.
\If {\textbf{not} $\hat{y}_{i-1}$ \textbf{and not} $\hat{y}_{i+1}$}~~ {// I-frame compression.}
        \State $\hat{y}_{\text{ref}} \Leftarrow$ None~~ {// Without reference feature.}
        \State $y_i^T \Leftarrow \mathcal{N}(0,1)$~~ {// Random noise.}
\ElsIf {$\hat{y}_{i-1}$ \textbf{and not} $\hat{y}_{i+1}$}~~ {// P-frame compression.}
        \State $\hat{y}_{\text{ref}} \Leftarrow \hat{y}_{i-1}$~~ {// Forward reference feature.}
        \State $y_i^T \Leftarrow$ DDIM\_Inversion $(\hat{y}_{\text{ref}}, T^{\prime}, \text{U-Net})$ ~~ {// Equation \ref{ddim} masked DDIM inversion.}
\ElsIf {$\hat{y}_{i-1}$ \textbf{and} $\hat{y}_{i+1}$}
        \State $\hat{y}_{\text{ref}} \Leftarrow O \cdot \hat{y}_{i-1}+(1-O) \cdot \hat{y}_{i+1}$~~ {// Bi-directional reference features.}
        \State $y_i^T \Leftarrow$ DDIM\_Inversion $(\hat{y}_{\text{ref}}, T^{\prime}, \text{U-Net})$~~ {// Eqution \ref{ddim} masked DDIM inversion.}
\EndIf 
\State {// Spatio-temporal variable-rate codec for controllable condition.}
\State $\hat{y}_i \Leftarrow D(\left\lfloor E(y_i, \hat{y}_{\text{ref}})\right\rceil, \hat{y}_{\text{ref}})$
\State {// Implicit inter-frame feature alignment using pre-trained U-Net.}
\For{$t = T ... 1$}
        \State $y_i^{t-1} \Leftarrow$ DDIM\_Backward $(y_i^t, t, \hat{y}_i, \text{U-Net})$~~ {// Transition kernel in Equation \ref{transition}.}
\EndFor
\State $\hat{x}_i=\mathcal{D}(y_i^0)$ // Frame reconstruction.
\State {\textbf{return}} $\hat{x}_{i},\hat{y}_i$
\end{algorithmic}
\end{algorithm}

\section{Methodology}
\subsection{Algorithm}
\label{al}
The algorithm of proposed I$^2$VC is illustrated in Section \ref{sec3}. The inputs are the target frame $x_i$ and the optional reference features $\hat{y}_{i-1}$ and $\hat{y}_{i+1}$ from the feature buffer. The outputs are the reconstructed frame $\hat{x}_i$ and the fusion feature $\hat{y}_i$.

\subsection{Implicit Inter-frame Feature Alignment}
\label{ad}

As described in Section~\ref{IIFA}, the state transition at each step $p_\theta(y_i^{t-1} {\mid} y_i^t)$ is considered as a motion compensation from the initial state $\hat{y}^{T^{\prime}}_\text{ref}$ towards the target feature $y_i$. The final denoised feature $y_i^0$ and the approximate motion $m_{\hat{y}_{\text{ref}} \rightarrow y_i}$ are formulated as:
\begin{equation}
\begin{aligned}
& y_i^0=\hat{y}_{\text{ref}}^{T^{\prime}}-\sum_1^T \varepsilon_\theta(y_i^t, t, \hat{y}_i)=\hat{y}_{\text{ref}}+\sum_1^{T^{\prime}} \varepsilon_\theta(\hat{y}_{\text{ref}}^t, t)-\sum_1^T \varepsilon_\theta(y_i^t, t, \hat{y}_i), \\
& m_{\hat{y}_{\text{ref}} \rightarrow y_i}\approx y_i^0-\hat{y}_{\text{ref}}=\sum_1^{T^{\prime}} \varepsilon_\theta(\hat{y}_{\text{ref}}^t, t)-\sum_1^T \varepsilon_\theta(y_i^t, t, \hat{y}_i),
\end{aligned}
\end{equation}
where $\varepsilon_\theta(\hat{y}_{\text{ref}}^t, t)$ and $\varepsilon_\theta(y_i^t, t, \hat{y}_i)$ represent the eliminated noise with U-Net. The motion $m_{\hat{y}_{\text{ref}} \rightarrow y_i}$ is estimated and compensated through the transition from the reference feature $\hat{y}_{\text{ref}}$ to the denoised feature $y_i^0$. The reference feature $\hat{y}_{\text{ref}}$ provides texture information and spatial structure for the denoised feature $y_i^0$, and the transition kernels provide implicit motion information. 

\begin{table}
\setlength\tabcolsep{9pt}
\renewcommand\arraystretch{1.3}
\centering
\begin{tabular}{ccccccc}
\hline
\toprule
\multicolumn{7}{c}{I$^2$VC-AI (I-frame training)}                                                              \\ \hline
Steps & Iterations & LR   & Frames   & I weight      & P weight        & B weight        \\ \hline
1    & 160K  & 1$e^{-4}$  & I                       & Pre-trained$\dag$                &              -                &         -                     \\
2    & 5K     & 1$e^{-5}$ & I                       & AI step 1$\dag$             &                  -            &          -                    \\ \hline
\multicolumn{7}{c}{I$^2$VC-LDP (P-frame training)}                                                                                                                    \\  \hline
1    & 160K  & 1$e^{-4}$  & I,P                    & AI step 2* & AI step 2$\dag$ &          -                    \\
2    & 5K    & 1$e^{-4}$  & I,P,P                 & AI step 2* & LDP step 1$\dag$ &          -                    \\
3    & 5K     & 1$e^{-5}$ & I,P,P                  & AI step 2* & LDP step 2$\dag$ &            -                  \\ 

\hline
\multicolumn{7}{c}{I$^2$VC-RA (B-frame training)}                                                            \\ \hline
1    & 5K  & 1$e^{-4}$  & I,B,I               & AI step 2* & -   & LDP step 3$\dag$ \\
2    & 5K    & 1$e^{-4}$  & I,B,B,B,I            & AI step 2* & -   & RA step 1$\dag$    \\
3    & 5K     & 1$e^{-5}$ & I,B,B,B,I           & AI step 2* & -   & RA step 2$\dag$    \\ 

\hline
\multicolumn{7}{c}{I$^2$VC-LDB (B-frame training)}               \\ \hline
1    & 5K  & 1$e^{-4}$  & I,P,B,P             & AI step 2* & LDP step 3*   & RA step 3$\dag$ \\
2    & 5K     & 1$e^{-5}$ & I,P,B,P             & AI step 2* & LDP step 3*   & LDB step 1$\dag$   \\ 
\hline
\toprule
\end{tabular}
\caption{The proposed training strategy for different compression configurations and frame types, where * denotes that the model parameters are loaded from the corresponding step and frozen in training, and $\dag$ denotes that only the U-Net is frozen.}
\label{train-str}
\end{table}

Consequently, it is crucial to use appropriate DDIM inversion steps $T'$ for implicit MEMC. The large number of steps implies that the target feature $y_i$ significantly diverges from the reference feature $\hat{y}_{\mathrm{ref}}$, posing a greater need for the motion compensation through controlled conditions $\hat{y}_i$. Conversely, the small step number indicates minimal temporal variations between the target feature $y_i$ and the reference feature $\hat{y}_{\text{ref}}$, allowing for greater use of the reference feature $\hat{y}_{\text{ref}}$ to reduce bit-rate. To standardize the number of DDIM inversion steps and enhance the model generalization, we set the number of steps as $T^{\prime}=\frac{1}{2} T$. 

\subsection{Training Strategy}
\label{ts}
Table \ref{train-str} suggests the different step-by-step training strategies for corresponding video compression configurations. For I-frame training, I$^2$VC-AI loads weights from the pre-trained LDM~\cite{rombach2022ldm} and first warms up the network by loss function $\mathcal{L}$ with 160K iterations with a Learning Rate (LR) of 1$e^{-4}$. I$^2$VC-AI is further fine-tuned through 5K iterations by a learning rate (LR) of 1$e^{-5}$. For P-frame training, I$^2$VC-LDP is initialized by I$^2$VC-AI, followed by a similar warm-up and fine-tuning steps, incorporating an additional step with an extra P-frame to assimilate inter-frame dependency. Thanks to the generalization of the unified framework I$^2$VC, the B-frame training of I$^2$VC-RA and I$^2$VC-LDB can be simplified as fine-tuning on I$^2$VC-LDP. All experiments are implemented on 4 NVIDIA GeForce RTX 3090 GPUs with Intel(R) Xeon(R) Gold 6248R CPUs. We conduct a mini-batch size of 4. The Adam optimizer~\cite{kingma2015adam} is utilized with $\beta_1 = 0.9$ and $\beta_2 = 0.999$. 

 \begin{figure}[!t]  
 \centering
  \begin{minipage}[b]{0.325\linewidth} 
   \centering
   {\includegraphics[width=\linewidth]{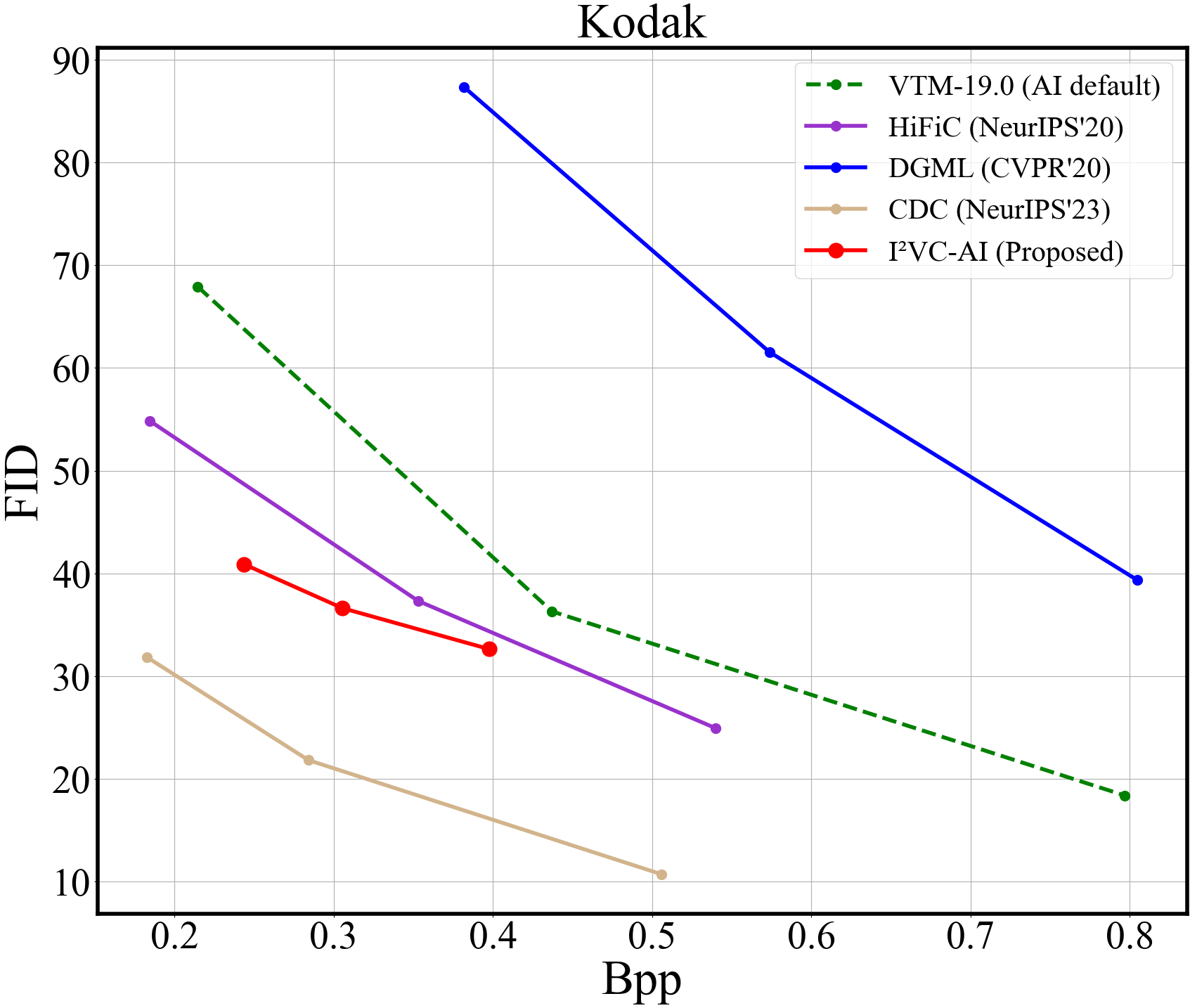}}
 \end{minipage}
 \hfill
 \begin{minipage}[b]{0.325\linewidth}
   \centering
   {\includegraphics[width=\linewidth]{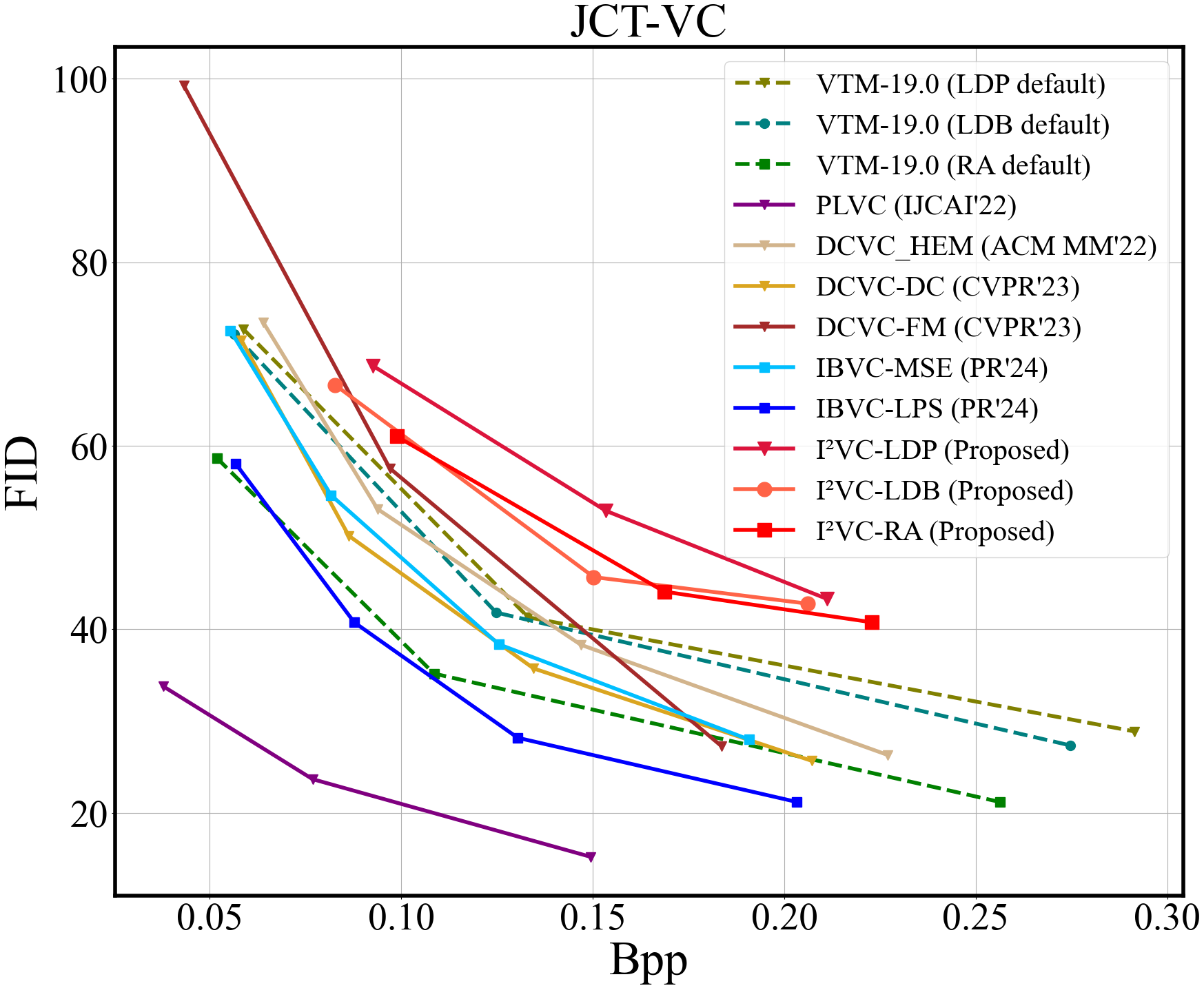}}
 \end{minipage}
 \hfill
 \begin{minipage}[b]{0.325\linewidth}
   \centering
   {\includegraphics[width=\linewidth]{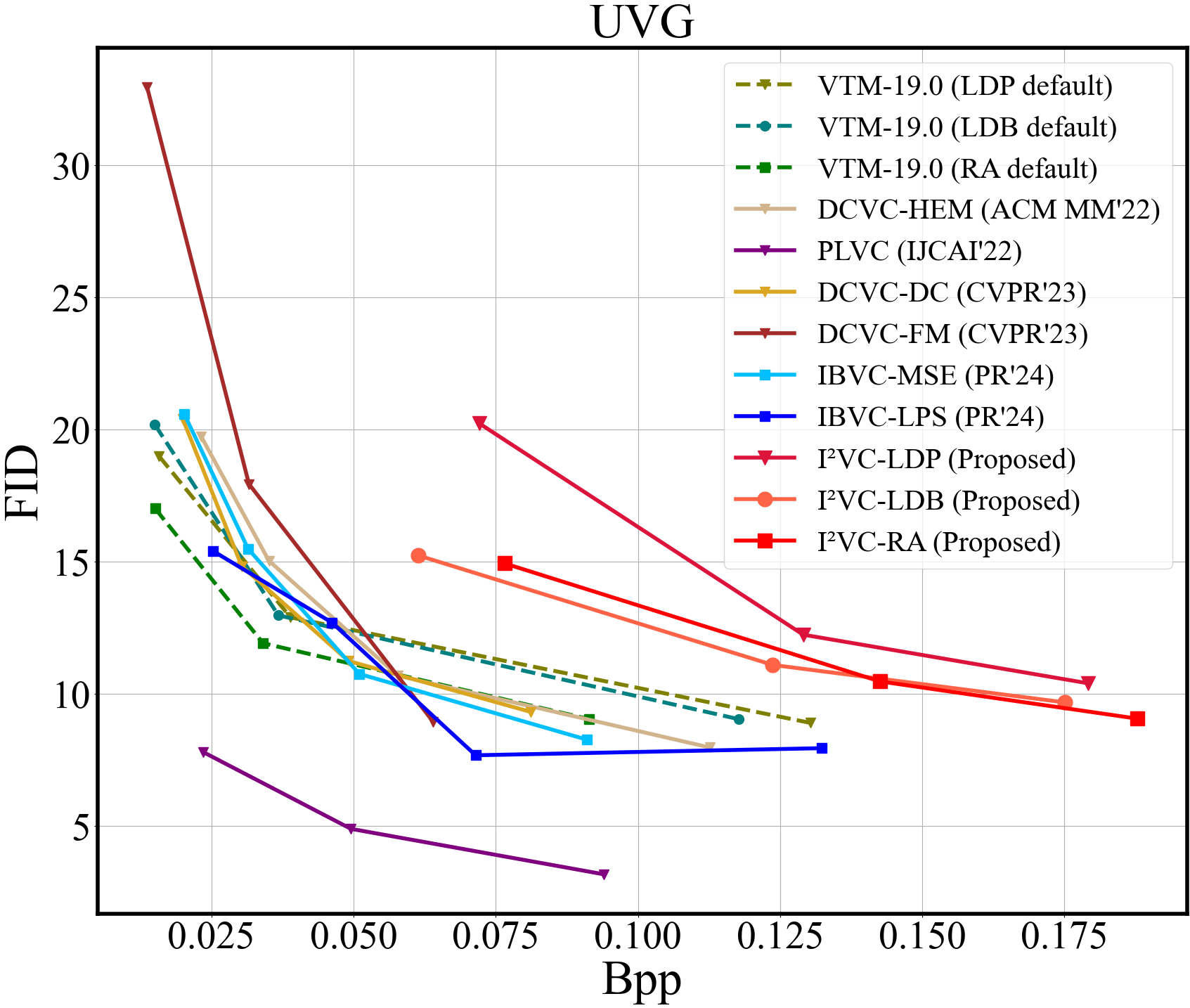}}
 \end{minipage}
   \caption{Rate-perception comparison of different video compression methods with three configurations on Kodak, JCT-VC and UVG datasets in terms of FID$\downarrow$.}
  \label{ap1}
\end{figure}

\begin{figure}[!t]
  \centering
  \includegraphics[width=\linewidth]{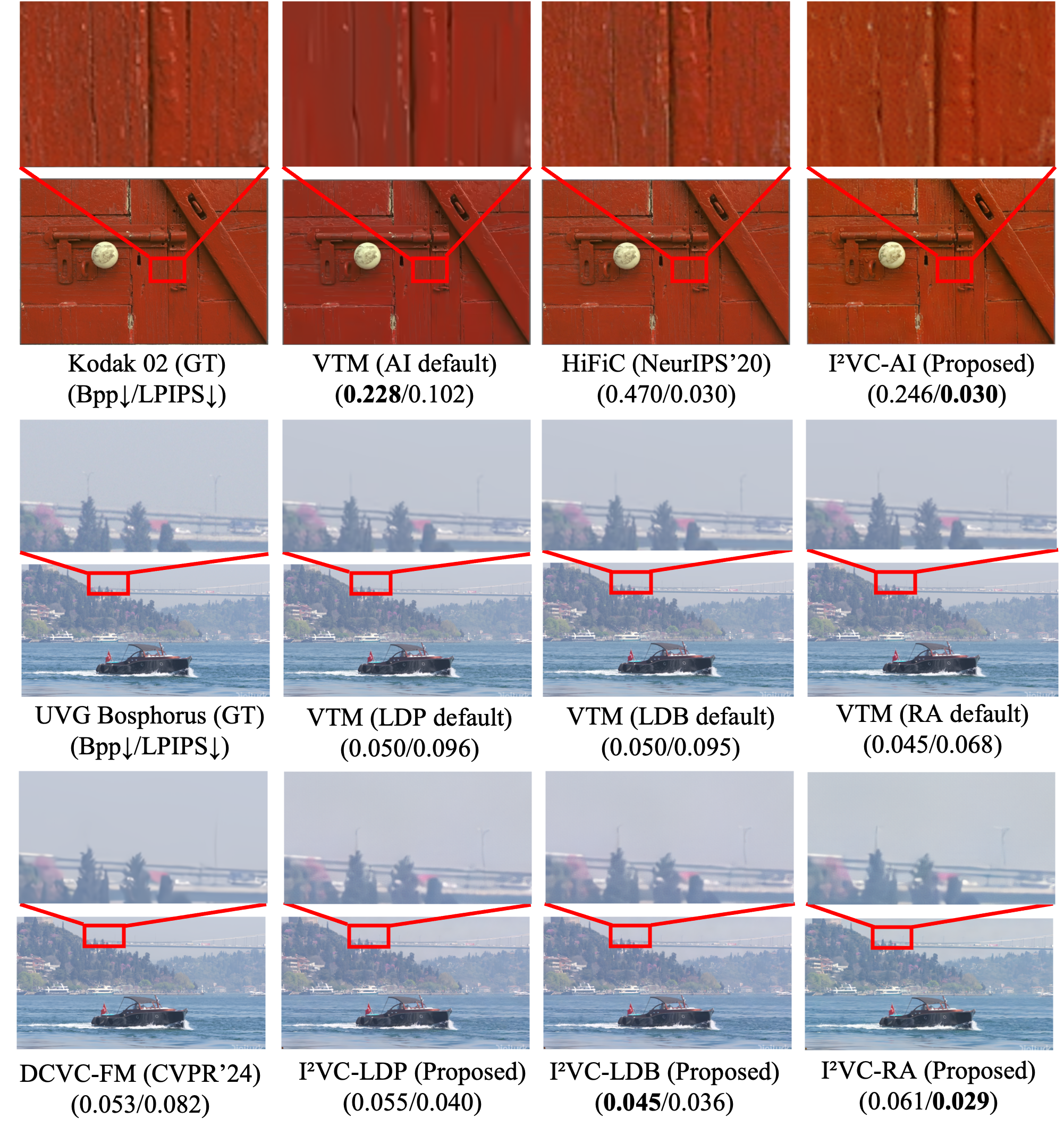}
  \caption{Qualitative comparison of different video compression configurations on Kodak and UVG datasets.}
  \label{ap2}
\end{figure}

\section{Experiments}
\subsection{Quantitative Evaluation}
\label{keguan}

As described in Figure \ref{ap1}, it is observed that I$^2$VC yields the acceptable rate-perception performance in terms of FID across three configurations, but does not achieve state-of-the-art performance like LPIPS metric. This is because the pre-trained LDM~\cite{rombach2022ldm} is trained using LPIPS loss, resulting in weak generalization on other metrics. However, the current FID performance still demonstrates the capability of I$^2$VC to consistently perform across the three configurations of video compression. Therefore, future work will focus on applying the training of diffusion models based on the findings of this study to address this limitation.

\subsection{Qualitative Evaluation}
\label{zhuguan}

As depicted in Figure \ref{ap2}. For AI configuration on Kodak dataset, I$^2$VC exhibits clearer wood texture with a lower bit-rate than the VTM-19.0 (AI default)~\cite{vtm} and HiFiC~\cite{mentzer2020high}. For LD and RA configuration on UVG dataset, I$^2$VC significantly reconstructs the street light on the left side. However, VTM-19.0~\cite{vtm} and DCVC-FM~\cite{li2024dcvcfm} with similar bit-rate of ours do not recover this street light. It proves that the implicit inter-frame alignment strategy of I$^2$VC can effectively maintain the invariant feature structure in video sequences and maintain temporal consistency.

% IMPORTANT, please:
% \begin{itemize}
%     \item {\bf Delete this instruction block, but keep the section heading ``NeurIPS paper checklist"},
%     \item  {\bf Keep the checklist subsection headings, questions/answers and guidelines below.}
%     \item {\bf Do not modify the questions and only use the provided macros for your answers}.
% \end{itemize} 

%%% END INSTRUCTIONS %%%

\end{document}